\documentclass[a4paper,10pt,twoside]{cpc-hepnp}
\usepackage{CJK,upgreek,fancyhdr}
\usepackage{multicol}
\usepackage{graphicx}
\usepackage{booktabs}
\usepackage{amssymb,bm,mathrsfs,bbm,amscd}
\usepackage[tbtags]{amsmath}
\usepackage{lastpage}

\usepackage[pagewise,columnwise]{lineno}
\begin{document}

\begin{CJK*}{GBK}{song}

\fancyhead[c]{\small Chinese Physics C~~~Vol. xx, No. x (202x) xxxxxx}
\fancyfoot[C]{\small 000-\thepage}

\footnotetext[0]{Received 10 October 2020}

\title{The observation of the Crab Nebula with LHAASO-KM2A for the performance study}

\author{
F. Aharonian$^{26,27}$,
Q. An$^{4,5}$,
Axikegu$^{20}$,
L.X. Bai$^{21}$,
Y.X. Bai$^{1,3}$,
Y.W. Bao$^{15}$,
D. Bastieri$^{10}$,
X.J. Bi$^{1,2,3}$,
\\
Y.J. Bi$^{1,3}$,
H. Cai$^{23}$
% ...
\\
(The LHAASO Collaboration)\email{chensz@ihep.ac.cn, caozh@ihep.ac.cn, wusha@ihep.ac.cn, licong@ihep.ac.cn, wangly@ihep.ac.cn, nanyc@ihep.ac.cn, lvhk@ihep.ac.cn, zhanghy607@mail.sdu.edu.cn, zhangyi@ihep.ac.cn}
}
\maketitle

\address{%
$^1$ Key Laboratory of Particle Astrophyics \& Experimental Physics Division \& Computing Center, Institute of High Energy Physics, Chinese Academy of Sciences, 100049 Beijing, China\\
$^2$University of Chinese Academy of Sciences, 100049 Beijing, China\\
$^3$TIANFU Cosmic Ray Research Center, Chengdu, Sichuan,  China\\
$^4$State Key Laboratory of Particle Detection and Electronics, China\\
$^5$University of Science and Technology of China, 230026 Hefei, Anhui, China\\
$^6$Department of Engineering Physics, Tsinghua University, 100084 Beijing, China\\
$^7$National Astronomical Observatories, Chinese Academy of Sciences, 100101 Beijing, China\\
$^8$National Space Science Center, Chinese Academy of Sciences, 100190 Beijing, China\\
$^9$School of Physics, Peking University, 100871 Beijing, China\\
$^{10}$Center for Astrophysics, Guangzhou University, 510006 Guangzhou, Guangdong, China\\
$^{11}$School of Physics and Astronomy \& School of Physics (Guangzhou), Sun Yat-sen University, 519000 Zhuhai, Guangdong, China\\
$^{12}$School of Physical Science and Technology, Guangxi University, 530004 Nanning, Guangxi, China\\
$^{13}$Hebei Normal University, 050024 Shijiazhuang, Hebei, China\\
$^{14}$School of Physics and Microelectronics, Zhengzhou University, 450001 Zhengzhou, Henan, China\\
$^{15}$School of Astronomy and Space Science, Nanjing University, 210023 Nanjing, Jiangsu, China\\
$^{16}$Key Laboratory of Dark Matter and Space Astronomy, Purple Mountain Observatory, Chinese Academy of Sciences, 210023 Nanjing, Jiangsu, China\\
$^{17}$Institute of Frontier and Interdisciplinary Science, Shandong University, 266237 Qingdao, Shandong, China\\
$^{18}$Key Laboratory for Research in Galaxies and Cosmology, Shanghai Astronomical Observatory, Chinese Academy of Sciences, 200030 Shanghai, China\\
$^{19}$Tsung-Dao Lee Institute \& School of Physics and Astronomy, Shanghai Jiao Tong University, 200240 Shanghai, China\\
$^{20}$School of Physical Science and Technology \&  School of Information Science and Technology, Southwest Jiaotong University, 610031 Chengdu, Sichuan, China\\
$^{21}$College of Physics, Sichuan University, 610065 Chengdu, Sichuan, China\\
$^{22}$Key Laboratory of Cosmic Rays (Tibet University), Ministry of Education, 850000 Lhasa, Tibet, China\\
$^{23}$School of Physics and Technology, Wuhan University, 430072 Wuhan, Hubei, China\\
$^{24}$School of Physics and Astronomy, Yunnan University, 650091 Kunming, Yunnan, China\\
$^{25}$Yunnan Observatories, Chinese Academy of Sciences, 650216 Kunming, Yunnan, China\\
$^{26}$Dublin Institute for Advanced Studies, 31 Fitzwilliam Place, 2 Dublin, Ireland \\
$^{27}$Max-Planck-Institut for Nuclear Physics, P.O. Box 103980, 69029  Heidelberg, Germany \\
$^{28}$ Dipartimento di Fisica dell'Universit\`a di Napoli   ``Federico II'', Complesso Universitario di Monte
                  Sant'Angelo, via Cinthia, 80126 Napoli, Italy. \\
$^{29}$Institute for Nuclear Research of Russian Academy of Sciences, 117312 Moscow, Russia\\
$^{30}$Moscow Institute of Physics and Technology, 141700 Moscow, Russia\\
$^{31}$D\'epartement de Physique Nucl\'eaire et Corpusculaire, Facult\'e de Sciences, Universit\'e de Gen\`eve, 24 Quai Ernest Ansermet, 1211 Geneva, Switzerland\\
$^{32}$Department of Physics, Faculty of Science, Mahidol University, 10400 Bangkok, Thailand\\
}

\begin{abstract}
As a sub-array of the Large High Altitude Air Shower Observatory (LHAASO), KM2A is mainly designed  to cover a large fraction of the northern sky to hunt for   gamma-ray sources  at energies above 10 TeV. Even though the detector construction is still underway, a half of the KM2A array has been   operating stably  since the end of 2019. In this paper, we present the pipeline of KM2A data analysis  and the first observation on the Crab Nebula, a standard candle in   very high energy gamma-ray astronomy.    We   detect  gamma-ray signals from the Crab Nebula in both energy ranges of 10$-$100 TeV and $>$100 TeV   with high significance, by analyzing the KM2A data of 136 live days between December 2019 and May 2020.
With the  observations, we   test  the  detector performance including angular resolution, pointing accuracy and cosmic ray background rejection power.
 The  energy spectrum of the Crab Nebula in the energy range 10-250 TeV fits well with a single   power-law function dN/dE =(1.13$\pm$0.05$_{stat}$$\pm$0.08$_{sys}$)$\times$10$^{-14}$$\cdot$(E/20TeV)$^{-3.09\pm0.06_{stat}\pm0.02_{sys}}$  cm$^{-2}$ s$^{-1}$ TeV$^{-1}$. It is consistent with previous measurements by other experiments.  This opens a new window of gamma-ray astronomy above 0.1 PeV through which    ultrahigh-energy gamma-ray new phenomena, such as cosmic PeVatrons,  might be discovered.
\end{abstract}

\begin{keyword}
Gamma-ray, Crab Nebula, Extensive air showers, Cosmic rays
\end{keyword}

\begin{pacs}
95.85.Pw, 96.50.sd, 98.70.Sa
\end{pacs}

\footnotetext[0]{\hspace*{-3mm}\raisebox{0.3ex}{$\scriptstyle\copyright$}
We are grateful for all the financial support of the   research work. Details are listed in the acknowledgement section.}%

\begin{multicols}{2}

\section{Introduction}
The Crab Nebula ($\sim$2 kpc  from the earth) is   the remnant of a core-collapse  supernova in 1054 AD recorded in Chinese and Japanese Chronicles \cite{green03}. Observations of the nebula have been carried out at every accessible wavelength, resulting in a remarkably well-determined spectral energy distribution (SED), making it  a ``standard candle" at several wavelengths up to very high energy (VHE). The Crab Nebula was the first VHE gamma-ray source discovered by the Whipple Collaboration in 1989 \cite{week89}. Up to now, the VHE emission has been firmly detected by many ground-based experiments, including both  air shower arrays \cite{bart13,bart15,abey19,amen19} and Imaging Cherenkov telescopes \cite{ahar04,ahar06,albe08}. Although several GeV flares  have been detected by AGILE and Fermi  \cite{tava11,abdo11}, the gamma-ray emission from the Crab Nebula is generally believed to be steady at higher energies. Recently, gamma-rays with energy above 100 TeV have been  detected by HAWC \cite{abey19} and Tibet AS$\gamma$ \cite{amen19}   from this source. The observed spectrum around 100 TeV is consistent with a smooth  extrapolation of the lower-energy   spectrum.  As a reference VHE gamma-ray source, the Crab Nebula is often used to check the detector performance, including sensitivity, pointing accuracy, angular resolution, and so on.

The non-thermal radiation of the Crab Nebula is characterized by  SED consisting of two  components.
The low-energy component extending from radio to gamma-ray frequencies comes from synchrotron radiation by relativistic electrons. The high-energy component   dominates the emission above $\sim$1   GeV and is produced  via inverse Compton (IC) scattering of ambient seed photons by relativistic electrons \cite{ahar04}. The absence of a high-energy cutoff in the measured spectrum from the Crab Nebula up to about 400 TeV  indicates that the primary electrons can reach at least  sub-PeV energies \cite{amen19}.

LHAASO (100.01$^{\circ}$E, 29.35$^{\circ}$N)  is a large hybrid extensive air shower (EAS) array being constructed at Haizi Mountain, Daocheng, Sichuan province, China \cite{he18}. It is composed of three sub-arrays, i.e., a 1.3 km$^2$ array (KM2A)  for gamma-ray astronomy above 10 TeV and cosmic ray physics, a 78,000 m$^2$ water Cherenkov detector array (WCDA)   for TeV gamma-ray astronomy, and 18 wide field-of-view air Cherenkov/fluorescence telescopes (WFCTA)  for cosmic ray physics from 10 TeV to  1 EeV. A considerable proportion of the LHAASO detectors have been operating since 2019 and the whole array will be completed in   2021.  KM2A has a wide field-of-view (FOV) of $\sim$2 sr   and covers 60\% of the sky within a diurnal observation. KM2A is unique for its unprecedented sensitivity at energy above 20 TeV. Even though only one half of KM2A has been operating for a few months, the   sensitivity for gamma-ray sources at energies above 50 TeV is already better than what has been achieved by  previous observations.

Here, we present the first observation of the ``standard candle" Crab Nebula using the first 5 months  half-array LHAASO-KM2A data from December 2019 to May 2020. Through this, the detector performance is thoroughly tested, including  pointing accuracy, angular resolution, background rejection power,  and flux determination.

\section{KM2A as an array for EAS detection}
\subsection{KM2A detector}
The whole KM2A array will consist  of 5195 electromagnetic detectors (EDs, 1 m$^2$ each) and 1188 muon detectors (MDs, 36 m$^2$ each),   deployed over an area of 1.3 km$^2$ as shown in Fig.~\ref{fig1}. Within 575 m from the center  of  the  array, EDs are distributed with a spacing of 15 m and MDs are distributed with a spacing of 30 m.  Within the outskirt ring region of width 60 m,  the spacing of ED   is enlarged to 30 m and these EDs are used to veto showers with cores located outside the central   1 km$^2$. KM2A    operates  around the clock   since both EDs and MDs can work during both  day and night.

An ED consists of 4 plastic scintillation tiles (100cm$\times$25cm$\times$1cm each). More details about the ED can be found elsewhere \cite{he18}. The coated tile is covered by a 5-mm-thick lead plate to absorb low-energy charged particles in showers and convert $\gamma$-rays into electron-positron pairs, which can improve the angular and core position resolution of the array. Once high energy charged particles enter   the scintillator, they  lose energy and excite the scintillation medium to produce a large amount of scintillation photons. The embedded wavelength-shifting fibers   collect scintillation light and transmit it to a 1.5-inch photomultiplier tube (PMT). The PMT  records the arriving time and number of the particles, based on which the  shower parameters can be reconstructed. The detection efficiency of a typical ED is about 98\%. The time resolution of an ED is about 2 ns. The resolution of the particle counter is $<$25\% for a single particle and the dynamic range is  from 1 to 10$^4$ particles. The average single rate of an ED is about 1.7 kHz with a threshold of 1/3 particle at the LHAASO site.

\begin{center}
\includegraphics[width=7cm]{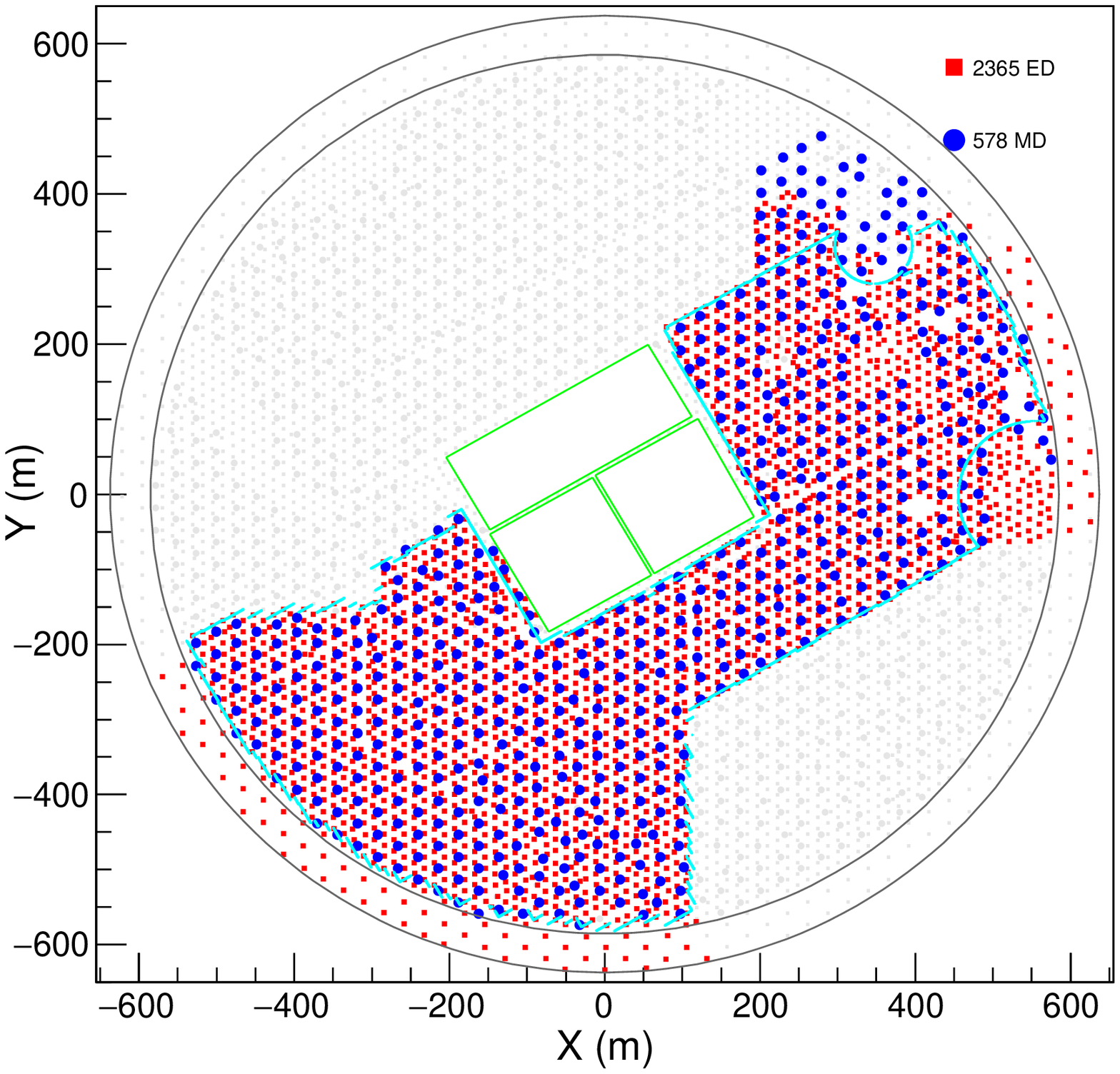}
\figcaption{\label{fig1} Planned layout of all LHAASO-KM2A detectors.
 The red squares  and blue circles indicate the EDs and MDs in operation, respectively.
 The area enclosed by the cyan   line outlines the fiducial area of the current  KM2A half-array used in this analysis.
 }
\end{center}

The MD is a pure water Cherenkov detector enclosed within a cylindrical concrete tank with an inner diameter of 6.8 m and height of 1.2 m. An 8-inch PMT is installed at the center of the top of the tank   to collect the Cherenkov light produced by high energy particles as they pass through the water. More details about the MD can be found elsewhere \cite{he18}. The whole detector is covered by a steel lid underneath soil.  The thickness of overburden soil is 2.5 m to   absorb  the secondary electrons/positrons and gamma-rays in showers. Thus the particles that   can reach the water inside and produce Cherenkov signals are almost exclusively muons, except for those MDs located at very central part of showers where some very high energy EM components may have a chance to punch through the screening soil layer.   The detection efficiency of a typical MD is $>$95\%.  The time resolution of an MD is about 10 ns. The resolution of the particle counter is $<$25\% for a single muon and the dynamic range is from 1 to 10$^4$ particles.  The average single rate of an MD is about 8 kHz with a threshold of 0.4 particles  at LHAASO site.

The detectors of KM2A were constructed and merged into the data acquisition system (DAQ) in stages. The first 33 EDs started  operating in February 2018 and the partial array was enlarged step by step afterward. Nearly half of the KM2A array, including 2365 EDs and 578 MDs and  covering
an area of 432 000  m$^2$  as shown in Fig.~\ref{fig1}, has been operating   since 27 December 2019.
The trigger logic of KM2A has been well tested and more details about it   can be found elsewhere \cite{wu18}. For the first half-array,  at least 20 EDs firing within a window of 400 ns is required for a shower trigger, thus yielding a negligible random noise trigger rate. The event trigger rate is about 1 kHz. For each event, the DAQ records 10 $\mu$s of data from all EDs and MDs that have   signals over the thresholds.

\begin{center}
\includegraphics[width=7cm]{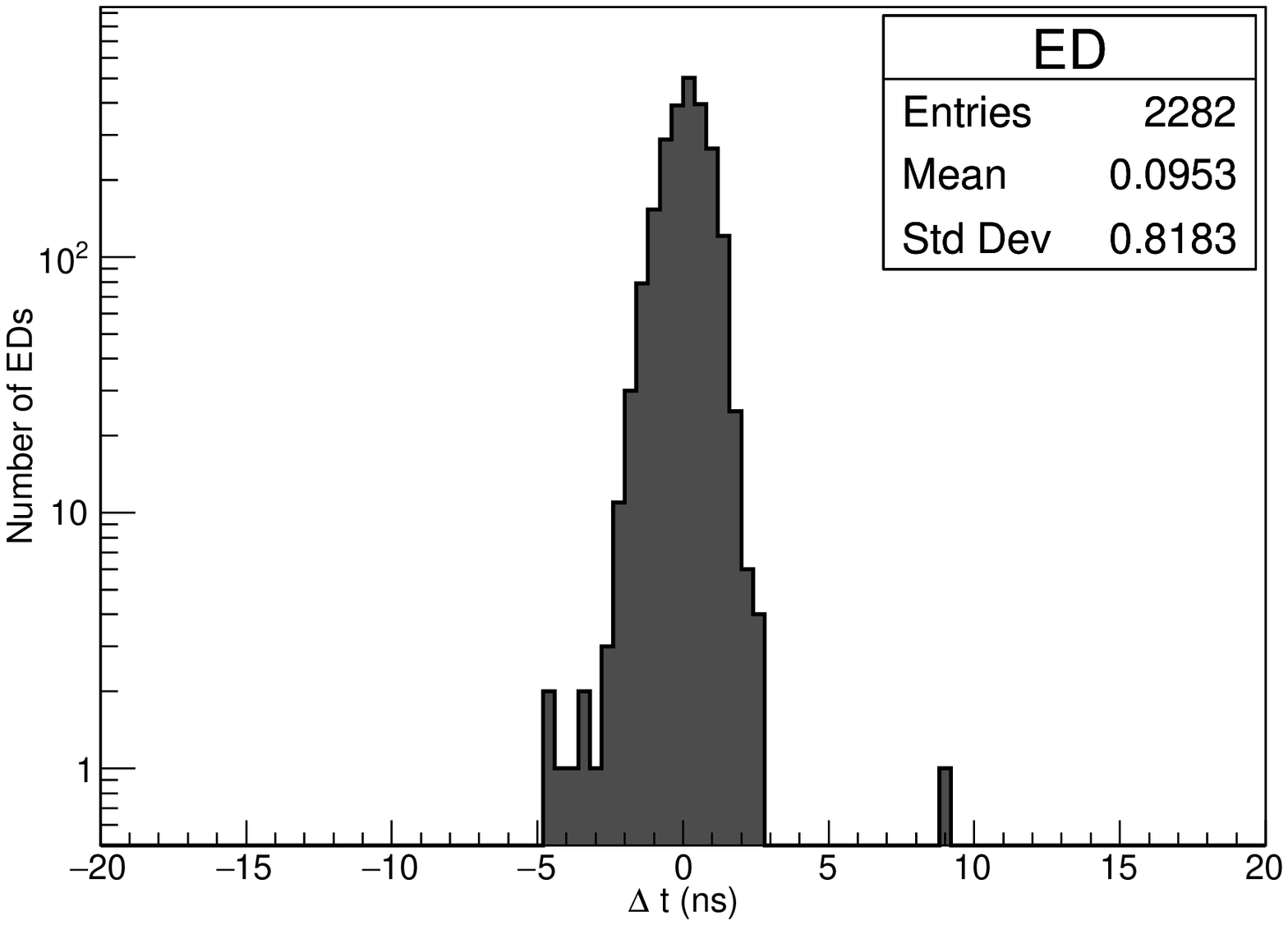}
\includegraphics[width=7cm]{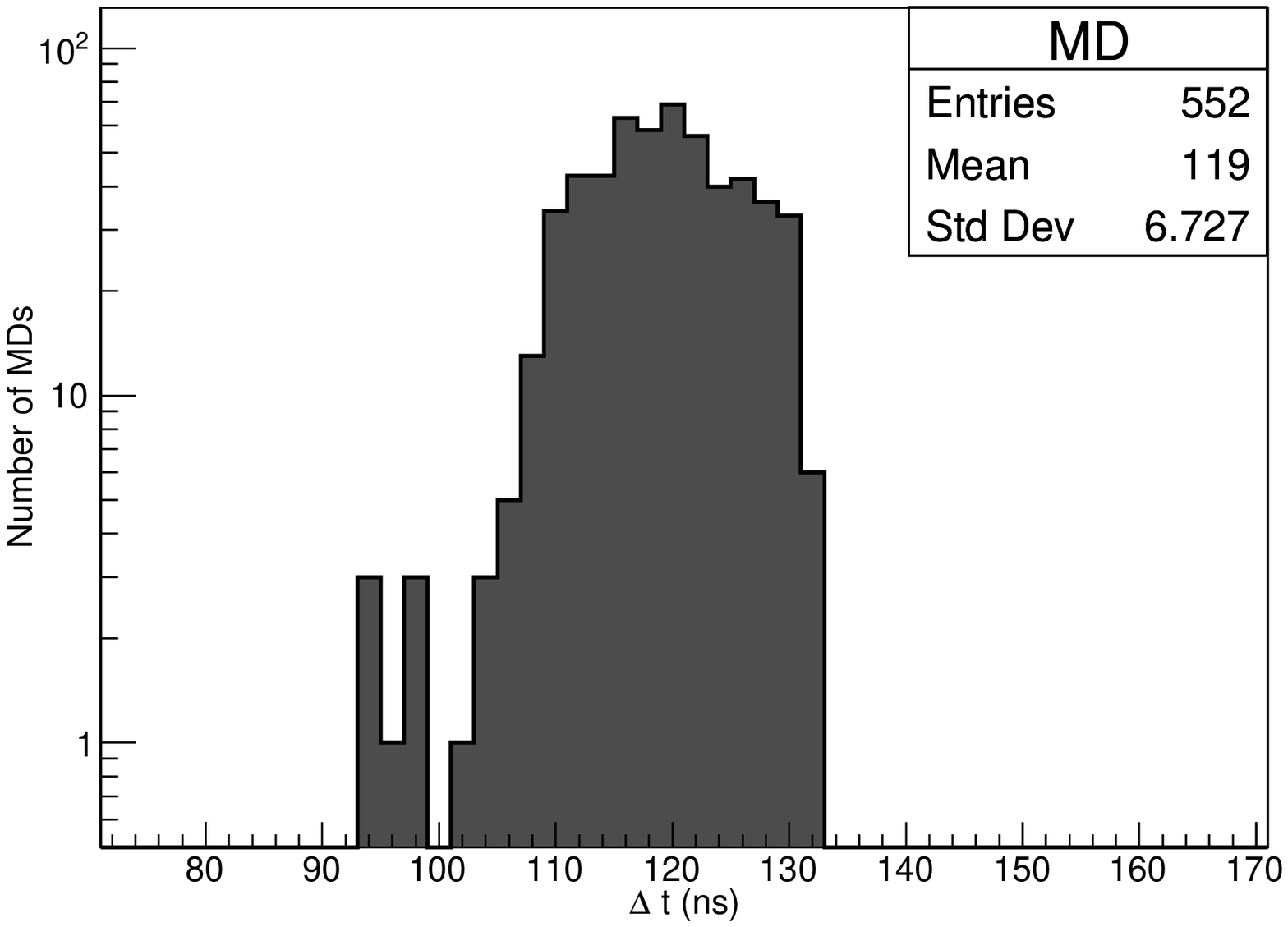}
\figcaption{\label{fig1a} The distribution of calibrated timing offset   for EDs and MDs using experimental data on 1st May 2020. The response of MDs has an average delay of 119 ns relative to that of EDs.
 }
\end{center}

The signal arrival time is measured by a time-to-digital converter (TDC) with a time precision  of 1 ns and 2 ns for EDs and MDs, respectively.  The clock of each TDC node is synchronized via the so-called White Rabbit (WR) timing system with an   accuracy of $\pm$150 ps. To further calibrate the detector response on timing measurement, an off-line method using the time residuals respect to a folded  shower front plane   is applied.  More details about this method can be found elsewhere \cite{lv18}.
Fig.~\ref{fig1a} shows the    calibrated timing offset   for EDs and MDs. The response of MDs averagely delays 119 ns relative to that of EDs.
The standard deviations are  0.82 ns among EDs and 6.7 ns among MDs, which can be taken to represent the timing uncertainty of individual detectors.
The timing calibration parameters are very stable and only need to be  updated about every one or two months.

The signal charge is measured by an analog-to-digital converter (ADC).  Based on the measurement of showers, the   typical charge produced by a particle for each ED and MD was calibrated.   With the charge calibration for each detector, the measured ADC counts were   converted into the number  of particles   \cite{lv18}. Affected  by the environment temperature, the calibration parameters for EDs and MDs  both vary with time.  The charge calibration parameters need to be  updated every day. A  variation about 5\% within each day remains uncorrected.

\subsection{Detector simulation}
To estimate properties of  primary  particles above the atmosphere, such as energy, composition, flux, etc, the simulation of detector response is crucial. In this work, the cascade processes within the atmosphere were simulated via the CORSIKA code (version7.6400) \cite{heck98}. To accurately simulate the KM2A detector response, a specific software G4KM2A  \cite{chen17,chen19} was developed in the framework of the Geant4 package (v4.10.00) \cite{agos03}. De-correlated single rate noise and corresponding charges determined by the experimental data are also taken into account in this simulation. This software adopts a flexible strategy and can simulate the KM2A array with any configuration. The reliability of the detector simulation was verified via the partial KM2A array data \citep{chen19}.
Fig.~\ref{fig1b} shows the distribution of  particle numbers   recorded by a typical   ED and MD, respectively. The simulation result is fairly consistent with experimental data.

In this work, a data sample with 2.222$\times$10$^8$ gamma-ray shower and 4.444$\times$10$^8$ proton shower events  was simulated. Both the gamma-ray and proton events are sampled in the energy range from 1 TeV to 10 PeV   following a power-law function with a spectral index of -2.0. The zenith angle is distributed from 0$^{\circ}$ to 70$^{\circ}$. The sample area is a circular region with a sufficiently large radius of 1000 m.

\begin{center}
\includegraphics[width=7cm]{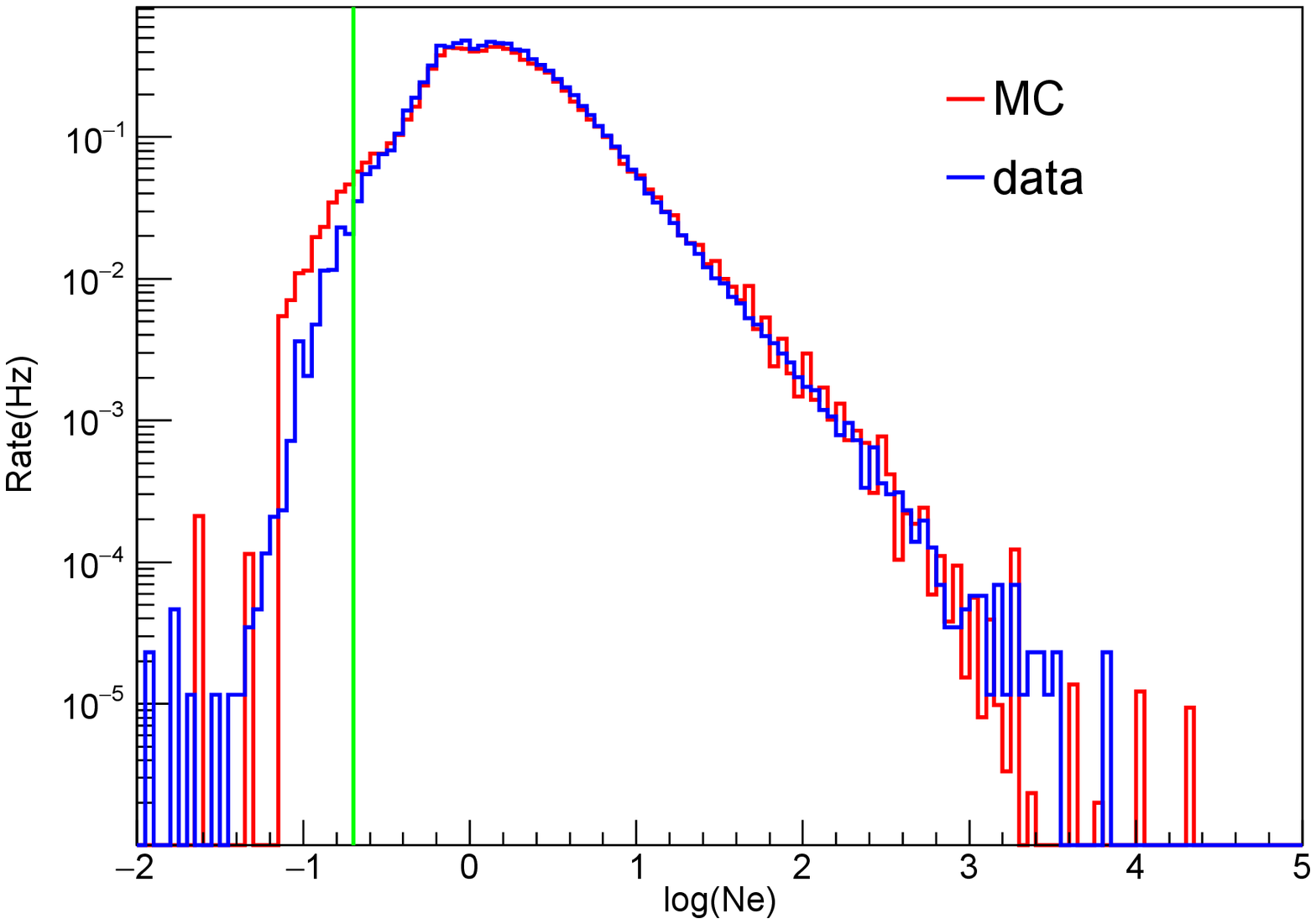}
\includegraphics[width=7cm]{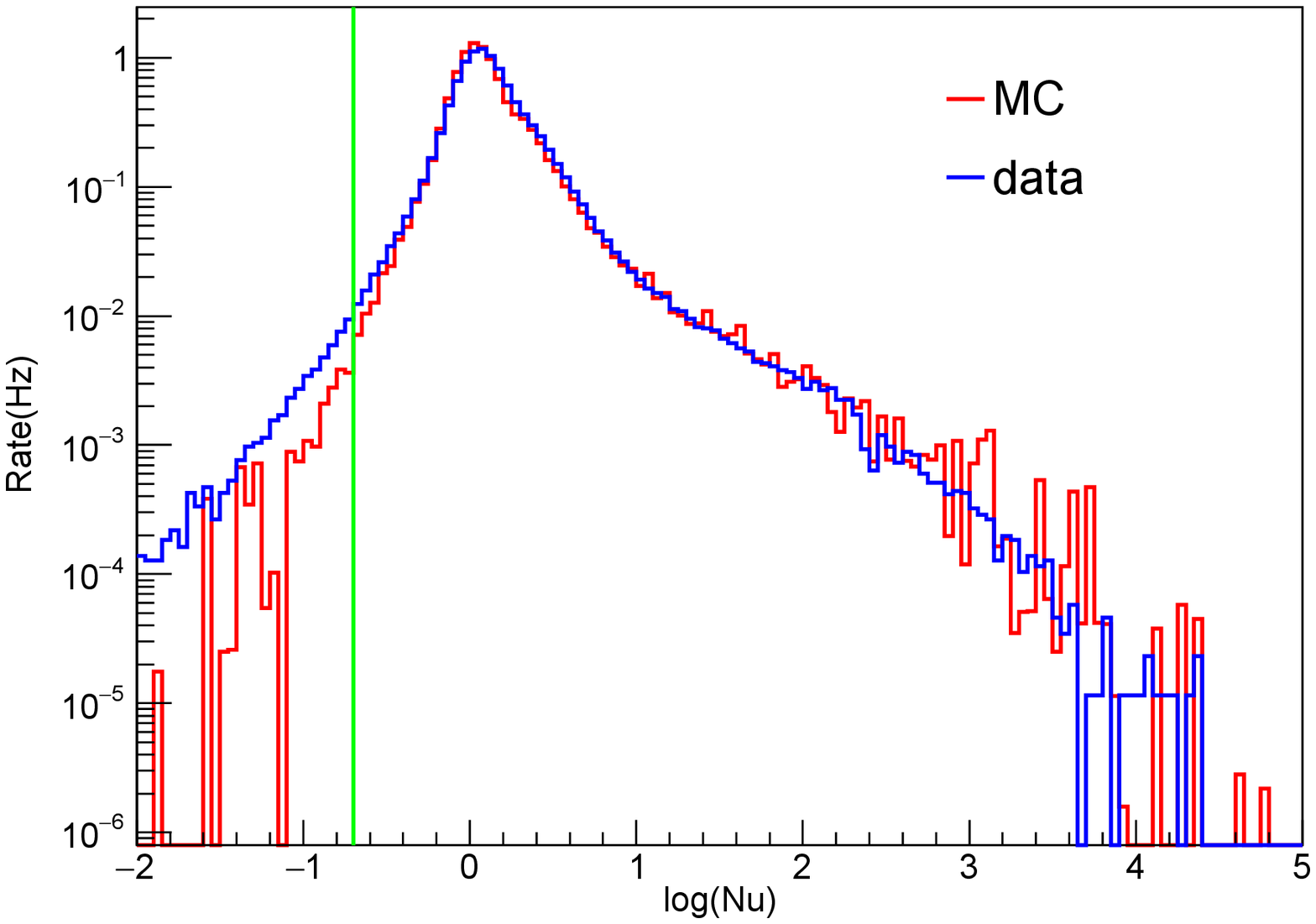}
\figcaption{\label{fig1b} The comparison between MC simulation  and experimental data of
the daily averaged trigger rate distribution of a typical ED (upper panel) and MD (lower panel). The horizontal axises indicate the number of particles recorded by these detectors for the triggered events.
The detectors with a particle number less than 0.2, as indicated by the vertical lines, are removed in both MC and experimental data reconstruction.
}
\end{center}

For a specific astrophysical source, the response of KM2A  depends both on the source emission spectrum and zenith angle within the detector FOV. With the above  data sample, we need re-normalize the distribution of the zenith angle for gamma-ray to  trace the  trajectory of the astrophysical source. In this work, the simulation data sample has been normalized to the sky trajectory of the Crab Nebula. In general the same data sample can be normalized for any astrophysical source of interest within the KM2A FOV. The response of KM2A  for a different emission spectrum can also be simulated via further normalization (weighting) on the primary spectrum. For the simulation data sample, the   data reconstruction pipeline for experimental data is adopted to extract the relevant quantities.

\section{Event reconstruction}
Each shower event is composed of many ED   and MD hits, each of which has  timing and charge information. In combination with the positions of these detectors, the primary direction and core location of the shower event can be reconstructed.  For KM2A events, only the ED hits are used for direction, core location, and energy reconstruction. Both ED hits and MD hits are used for composition discrimination. For the experimental data selection, first the status of  each detector is evaluated and    abnormal EDs and MDs are removed. Then,  each hit is calibrated for its timing and charge information to unify the detector response.   Finally, both experimental and simulation events  are processed  through the same reconstruction pipeline.

For the event reconstruction, firstly, a time window of 400ns and a circular   window with a radius of 100 m are adopted to select the most probable real secondary shower hits. With these selected hits, the core location is reconstructed using an optimized centroid method and the direction is reconstructed by fitting the shower plane. Secondly, only hits  within [-30,50] ns of the shower plane and   with a distance less than 200 m from the shower core are selected. Using these  hits, the core location,   shower size (denoted as $N_{\rm size}$) and age (denoted as s) are   reconstructed using a likelihood method.    The direction is also updated.  Finally, all the ED   and MD hits within [-30,50] ns of the reconstructed shower plane are selected. The final surviving ED hits   are used to count the number of electromagnetic particles (denoted as N$_{\rm e}$).   To reduce  pollution from the punch-through   high energy electromagnetic particles near the shower core, only   MDs with a distance farther than 15 m from the core are used to obtain the number of muons N$_{\mu}$. The parameters N$_{\mu}$ and N$_{\rm e}$ are used to discriminate between gamma-ray showers and  cosmic ray showers.

We show in Fig.~\ref{fig2}  the pattern of a high energy gamma-ray like shower ( N$_{\mu}$=0 )  detected by KM2A from the Crab Nebula direction.
Although there are many random noise hits during the recorded shower, the core location is evident from the distribution of particle density. The particle density and arrival time of the shower become   clear after filtering out the noise hits via reconstruction.

Before giving details  about the core location, direction, energy reconstruction, effective area, and gamma-ray/backgroud discrimination we list several  data quality cuts: (1) shower core is located in the fiducial area enclosed by the
cyan lines in Fig. 1; (2) the zenith angle is less than 50$^{\circ}$; (3) the number of particles detected within 40m from shower core is larger than that within 40$-$100m.
(4) the number of EDs and the number of particles for the reconstruction are both greater than  10. (5) the shower age is between 0.6 and 2.4.

\end{multicols}
\begin{center}
\includegraphics[width=5.3cm,height=5.3cm]{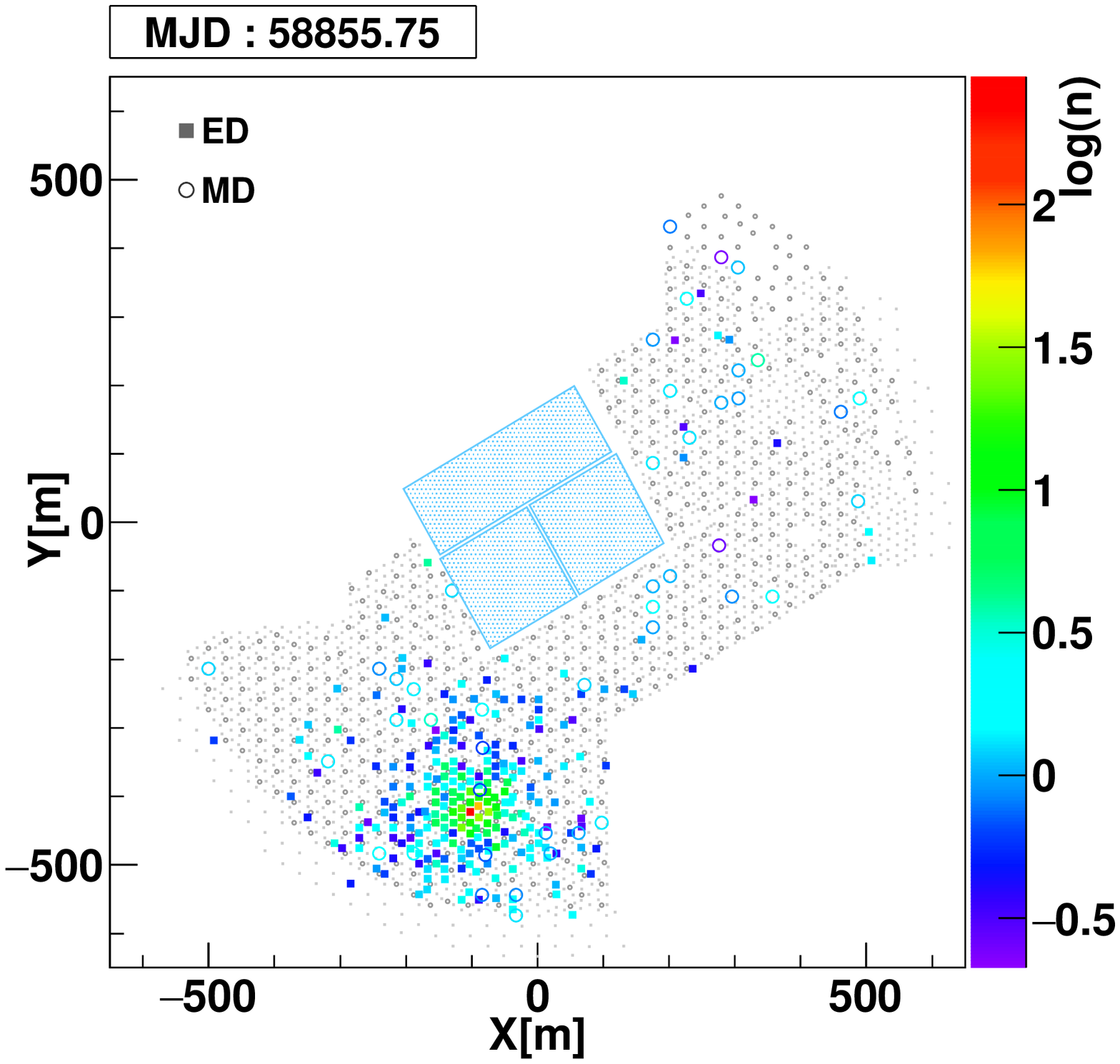}
\includegraphics[width=5.3cm,height=5.3cm]{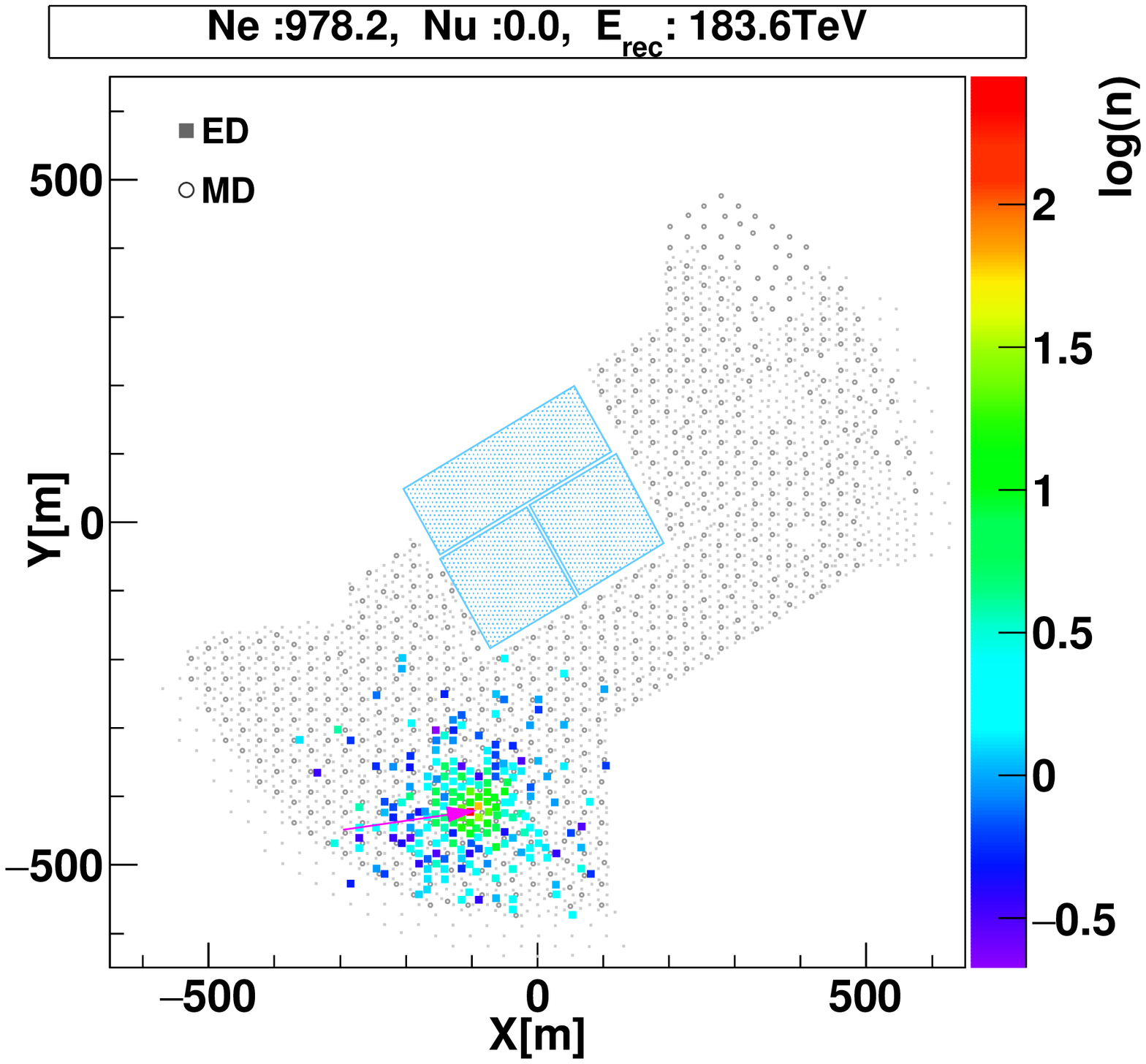}
\includegraphics[width=5.3cm,height=5.3cm]{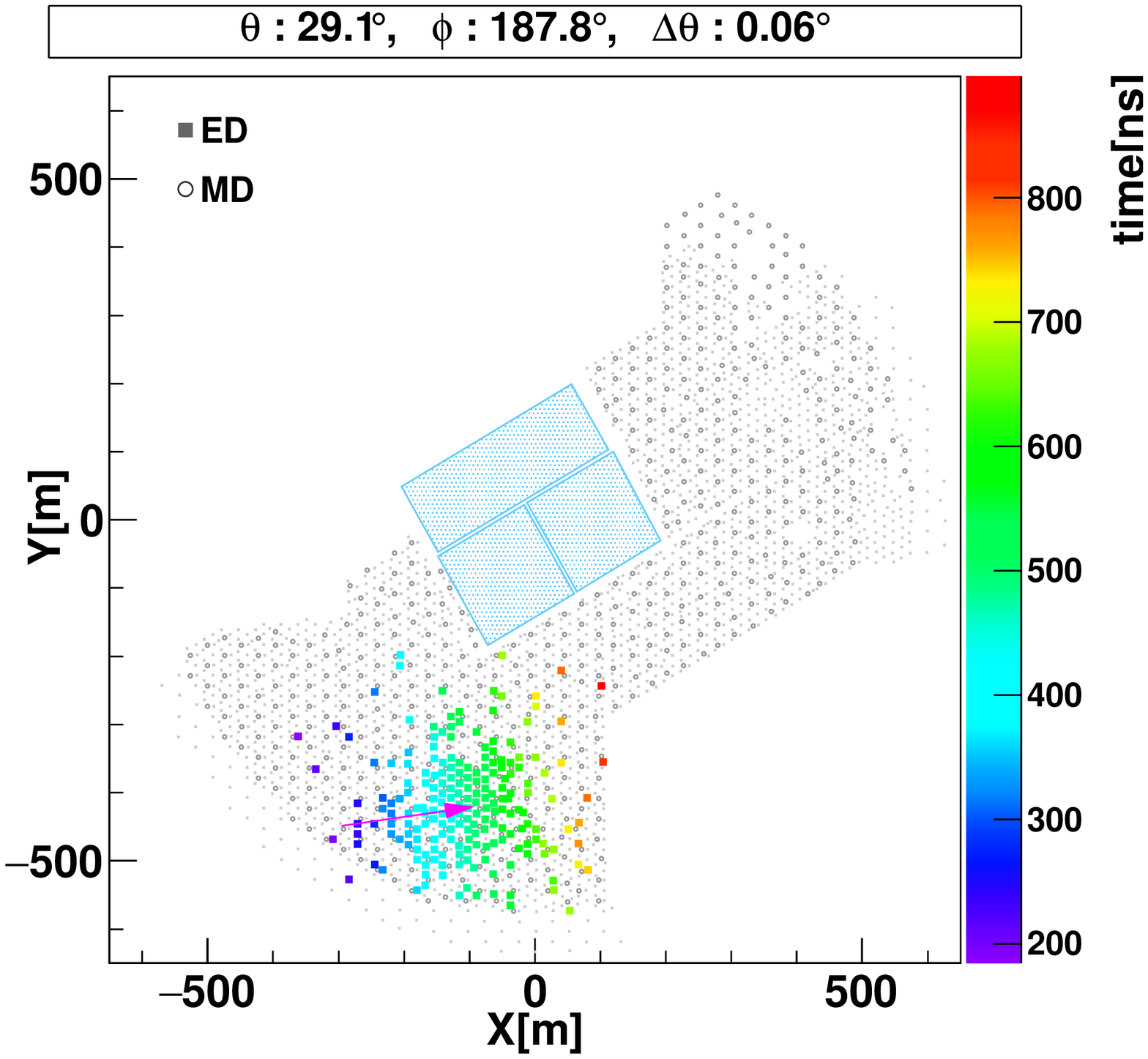}
\figcaption{\label{fig2} A high-energy gamma-ray-like shower detected by KM2A from the Crab Nebula. Left: the original particle density of detector units in KM2A. Middle: the  particle density map after filtering out noise hits  that are clearly irrelevant to the reconstructed shower front. The color scale indicates the logarithm of the particle density.  Right: the unit map of the arrival time.  The color scale indicates the relative trigger time of the unit in $ns$. E$_{rec}$ denotes the reconstructed energy of the event. $\theta$ and $\phi$ denote the zenith angle and azimuth angle of the event, respectively.  The red arrows shows the incident direction  of the event.
$\Delta\theta$  denotes the space angle between  this event and the   Crab Nebula direction.
}
\end{center}
\begin{multicols}{2}

\subsection{Core reconstruction}
In an air shower, most of the secondary particles are distributed along the trajectory of the original primary particle. The expected position of primary particle on the ground is defined as the shower core. Determining   the core location is crucial for direction reconstruction, which will use the core location as a fixed vertex when fitting the shower front to a conical shape. The simplest method to reconstruct the core position consists of calculating the average of the fired detector coordinates weighted with the number of particles (denoted as $n_{\rm e}$). This simple algorithm is called the centroid method which is fast in   computing  time  while it turns out to be inadequate to perform a good core reconstruction. More refined techniques are needed.
In this work, an optimized centroid method is  implemented first. The functions are:
\begin{equation}
Corex=\frac{\sum w_{\rm i}   x_{\rm i}}{\sum w_{\rm i}}, Corey=\frac{\sum w_{\rm i}   y_{\rm i}}{\sum w_{\rm i}},  Corez = \frac{\sum w_{\rm i}   z_{\rm i}}{\sum w_{\rm i}},
\end{equation}
where $w_{\rm i}$=$n_{e}e^{-0.5(r_{\rm i}/15)^2}$, (x$_{\rm i}$, y$_{\rm i}$, z$_{\rm i}$) are the ED coordinates, and
$r_{\rm i}$ is the ED distance to the shower core and  the unit is m. The calculation   needs about 20 iterations before converging.
The obtained core location is used to filter out noise hits and  as initial values for further core reconstruction.

The core is further reconstructed by fitting   the lateral distribution  function of the shower.
The lateral distribution of particle density measured by the KM2A array is fitted using the
following modified Nishimura-Kamata-Greisen (NKG) function \cite{greis60}:
\begin{equation}
\rho(r)=\frac{N_{\rm size}}{2 \pi r_{\rm m}^{2}}\frac{\Gamma(4.5-s)}{\Gamma(s-0.5)\Gamma(5-2s)}(\frac{r}{r_{\rm m}})^{s-2.5}(1+\frac{r}{r_{\rm m}})^{s-4.5}
\end{equation}
where r is the distance to the air shower axis, $N_{\rm size}$ is the total number of particles,  s is the age of the shower, and r$_{\rm m}$ is Moli\`ere radius.
r$_{\rm m}$ is  fixed at 136 m. The reconstructed parameters are the core location, $N_{\rm  size}$ and  s.
The MINUIT package \cite{james75} is used to maximize the log likelihood by varying the parameters via two steps.
Firstly, the core location is reconstructed with   s=1.2. Secondly,   $N_{\rm size}$ and  s are reconstructed with the  core location fixed at values obtained from the first step.
 Fig.~\ref{fig2b} shows the lateral distribution and corresponding fitting result of the gamma-ray-like event shown in Fig.~\ref{fig2}.

\begin{center}
\includegraphics[width=7.cm,height=5.cm]{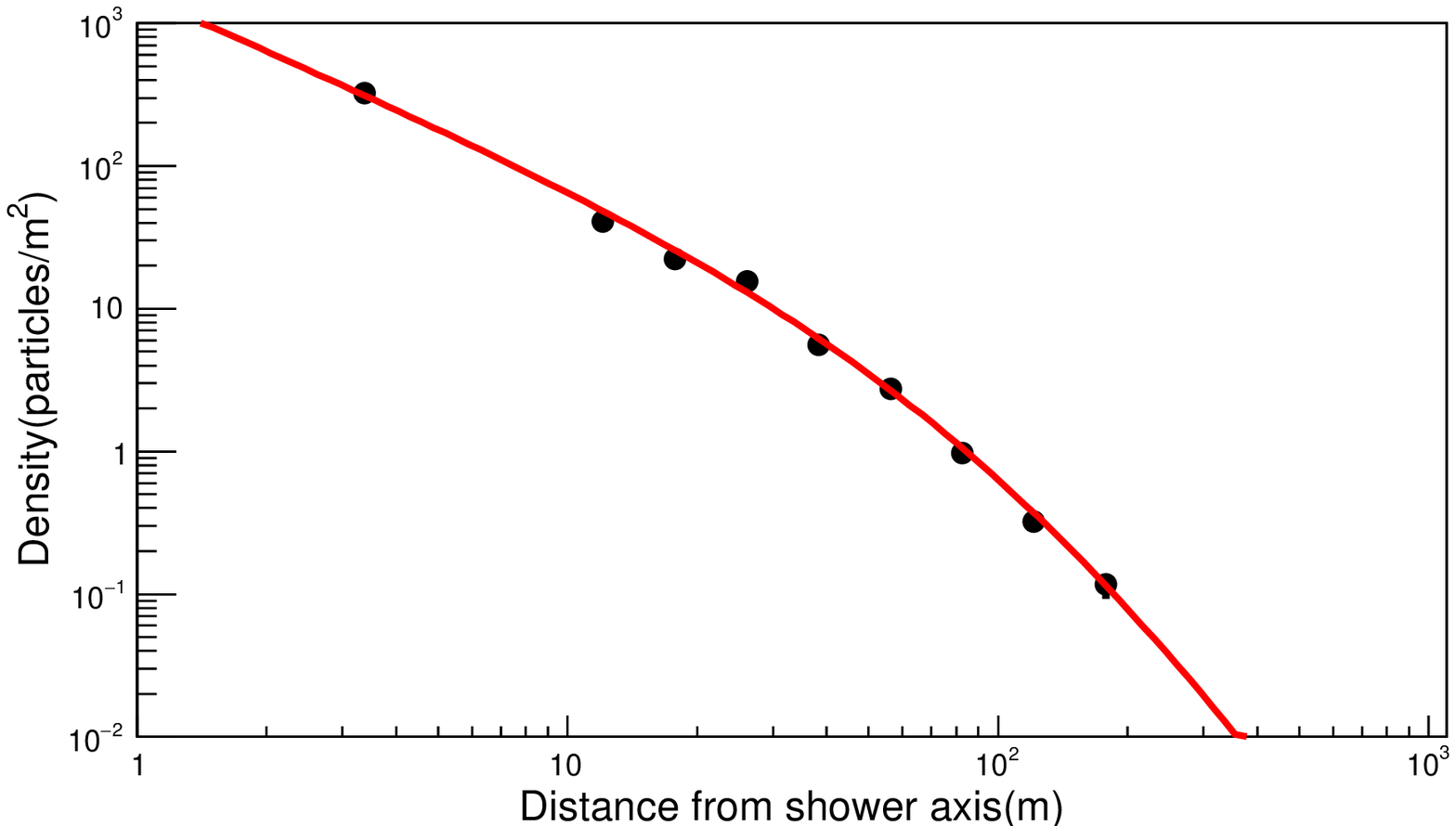}
\figcaption{\label{fig2b} The lateral distribution of the gamma-ray-like event shown in Fig.~\ref{fig2}. The solid curve shows the modified NKG function (2) that   fits   the data. The energy is  184$\pm$31 TeV.
 }
 \end{center}

The core resolution is  energy and zenith dependent.  The core resolution for gamma-ray events over various zenith angle ranges  is shown in Fig.~\ref{fig5} as a function of the reconstructed energy.  The resolution (denoted as R$_{\rm 68}$, containing 68\% of the events) is about 4$-$9 m at 20 TeV and 2$-$4 m at 100 TeV.

\begin{center}
\includegraphics[width=7cm]{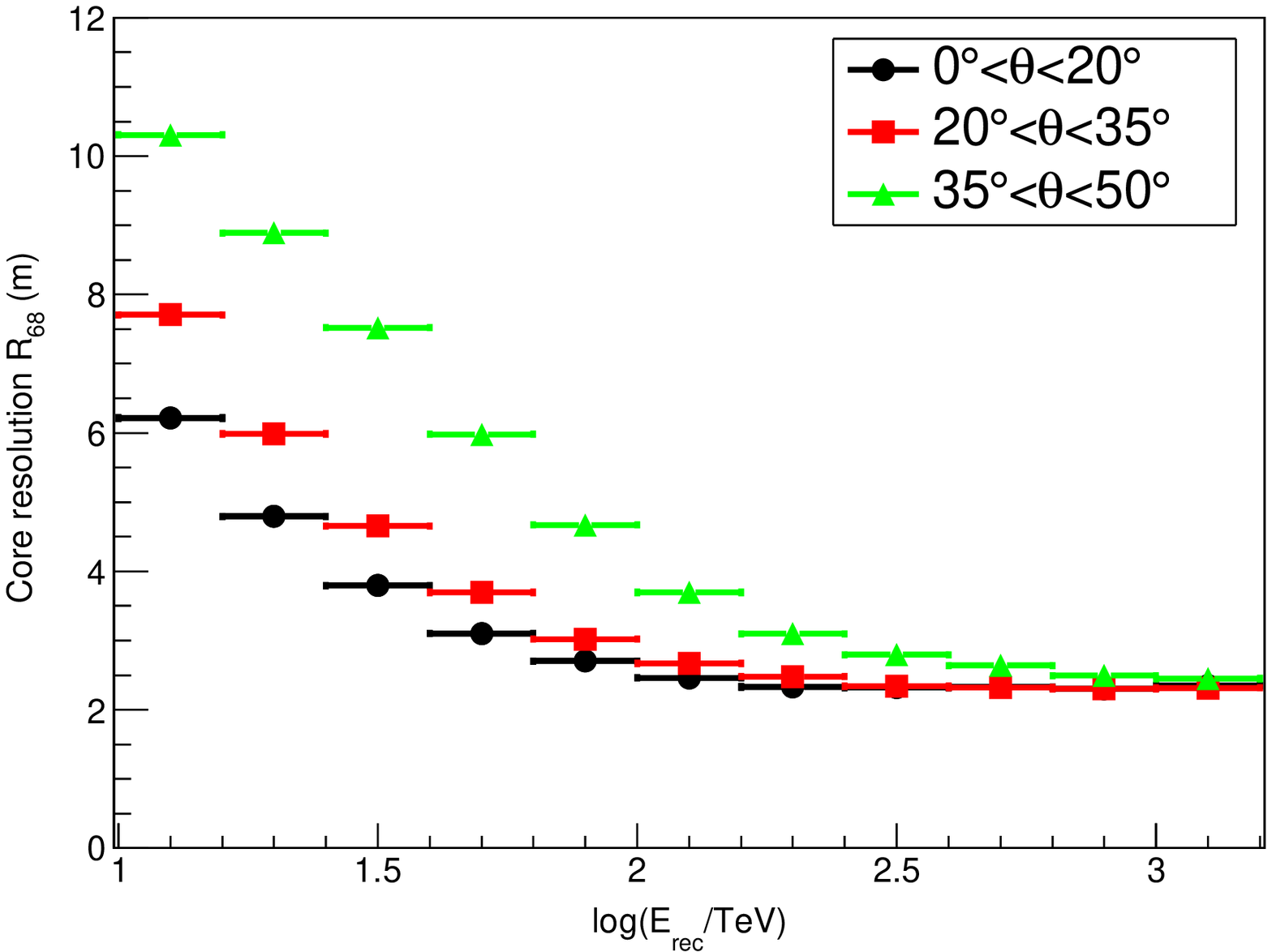}
\figcaption{\label{fig5}Core resolution of the KM2A half-array for simulated gamma-ray showers over different zenith angle ranges.}
\end{center}

\subsection{Direction reconstruction}
The secondary particles of a shower roughly travel in a plane perpendicular to the direction of the primary particle with the speed of light, as illustrated in Fig.~\ref{fig2}. The direction can be reconstructed by fitting the shower plane. In fact, the shower front has a slightly conical shape which needs to be accounted for when performing a good direction reconstruction. The arrival times of the particles are fitted by minimizing the following quantity:
\begin{equation}
\chi^{2}=\frac{1}{N_{hit}}\sum_{\rm i=1}^{\rm i=N_{\rm hit}} w_i(t_{\rm i}-l\frac{x_{\rm i}}{c}-m\frac{y_{\rm i}}{c}-n\frac{z_{\rm i}}{c}-\alpha\frac{r_{\rm i}}{c}-t_{\rm 0})
\end{equation}
where $l$=sin$\theta$cos$\phi$, $m$=sin$\theta$sin$\phi$, $n$=cos$\theta$,   $\theta$ and $\phi$ are the direction angles, $\alpha$ is the conical correction coefficient, and c=0.2998 m/ns is the speed of light.
The sum is over the fired EDs, t$_{\rm i}$ is the measured time of the $i$th ED, x$_{\rm i}$, y$_{\rm i}$, z$_{\rm i}$ are the ED coordinates,
 r$_{\rm i}$ is the ED distance from the  core in the shower plane, and $w_i$ is a weight set according to  the time residual and distance to the shower core, i.e.,  $w = \zeta(\delta t )\cdot \xi(r)$.
It is  known that the distribution of time residuals relative to the shower front is asymmetric. The multiple scattering can lead to a broader arrival time distribution for  delayed particles. To optimize the fit, a specific asymmetric weight method is adopted in this work to reduce the effect of the delayed particles.  The weight is set according to:
\begin{equation}
\begin{aligned}
\zeta(\delta t )=
\begin{cases}
1,  \qquad \qquad \qquad (-20 < \delta t < 0) \\
e^{-\frac{1}{2}(\frac{ \delta t}{10})^{2}},  \qquad  ( \delta t>0, \delta t < -20 )
\end{cases}  \\
\delta t=t_{\rm i}-l\frac{x_{\rm i}}{c}-m\frac{y_{\rm i}}{c}-n\frac{z_{\rm i}}{c}-\alpha\frac{r_{\rm i}}{c}-t_{0} \qquad \qquad \\
\end{aligned}
\end{equation}
where the times are given in ns.

It is also known that the error in the arrival time  increases with the distance from the shower core. An empirical function in \cite{linsl85} is   used in this work to calculate the weight according to the distance from the shower core. The weight is
\begin{equation}
\begin{aligned}
\xi(r)=\frac{1}{\sqrt{1+(1.6(\frac{r}{30}+1)^{1.5})^{2}}}
\end{aligned}
\end{equation}
where r is given in m.

The reconstructed parameters are the direction cosines
$l$, $m$,  $\alpha$ and the offset time t$_0$. $n$ can be determined using the parameters $l$ and $m$ during iteration.  The zenith angle $\theta$ and azimuth angle $\phi$ of the shower can be derived from the parameters $l$ and $m$.
The angular resolution is  energy and zenith angle dependent.  The angular resolution for gamma-ray events  is shown in Fig.~\ref{fig6} as a function of the reconstructed energy over different zenith angle ranges.
The resolution (denoted as $\phi$$_{68}$,  containing 68\% of the events) is  0.5$^{\circ}-$0.8$^{\circ}$ at 20 TeV and 0.24$^{\circ}-$ 0.3$^{\circ}$  at 100 TeV.

\begin{center}
\includegraphics[width=7cm]{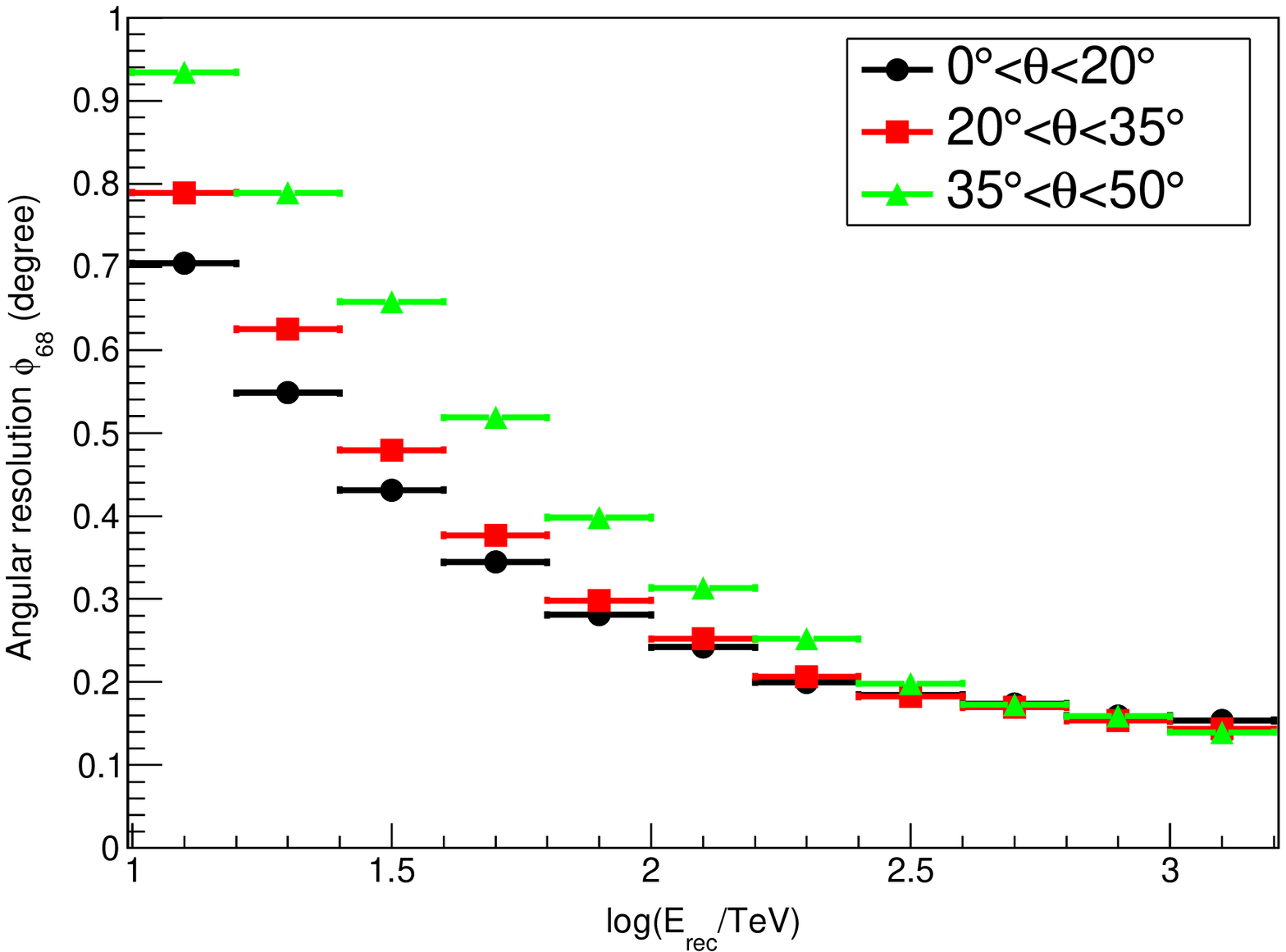}
\figcaption{\label{fig6}Angular resolution of the KM2A half-array for simulated  gamma-ray showers over different zenith angle ranges.}
\end{center}

\subsection{Energy reconstruction}
EAS arrays work by detecting the shower particles that reach ground level.
A simple way to estimate a shower energy  is to count the number of triggered detector elements, as
  used by the ARGO-YBJ experiment \cite{bart18}. A robust estimator of a shower energy is to utilize the normalization of  the lateral distribution function (LDF) of the shower as  proposed by \cite{hilla71}. Usually, this is implemented by using the particle density at the optimal radius  at which the uncertainty  is minimized.
 This method has been  used by  Tibet AS$\gamma$ \cite{kawat17} and HAWC \cite{abey19}.

The particle density at r=50m (denoted as $\rho_{50}$) evaluated using Equation (2) is used to estimate the gamma-ray energy in this work. The energy resolution values using densities from $\rho_{40}$ to $\rho_{70}$ are
almost the same.
Because the atmospheric depth over which the shower develops is proportional to sec($\theta$), the zenith angle effect has to be taken into account in the energy reconstruction. The final response function between $\rho_{50}$ and the primary energy is given by:
\begin{equation}
log(E_{\rm rec}/TeV)=a(\theta)\cdot (log(\rho_{50}))^{2} +b(\theta)\cdot log(\rho_{50}) +c(\theta)
\end{equation}
where $a(\theta)$, $b(\theta)$  and $c(\theta)$ are known constants, which have been given as functions of    sec($\theta$).
The shower illustrated in Fig.~\ref{fig2} is estimated to have energy 184$\pm$31 TeV using equation (6).

The energy resolution is energy and zenith angle dependent.
Fig.~\ref{fig3} shows the relation between the reconstructed energy (E$_{\rm rec}$) and the primary true energy (E$_{\rm true}$) over zenith angles   0$^{\circ}$-50$^{\circ}$. As the energy of the primary gamma-ray increases, the shower maximum becomes closer
to the altitude of the observatory,  leading to better energy resolution. As the zenith angle increases, the shower maximum becomes higher,  leading to a worse energy resolution.
In this work, the events with reconstructed energy above 10 TeV   are divided into  five bins per decade.
The energy resolution for each energy bin over different zenith angles is shown in Fig.~\ref{fig3}. For showers with zenith angle less 20$^{\circ}$, the resolution is about 24\% at 20 TeV and 13\% at 100 TeV.

\end{multicols}
\begin{center}
\includegraphics[width=5.3cm,height=5.3cm]{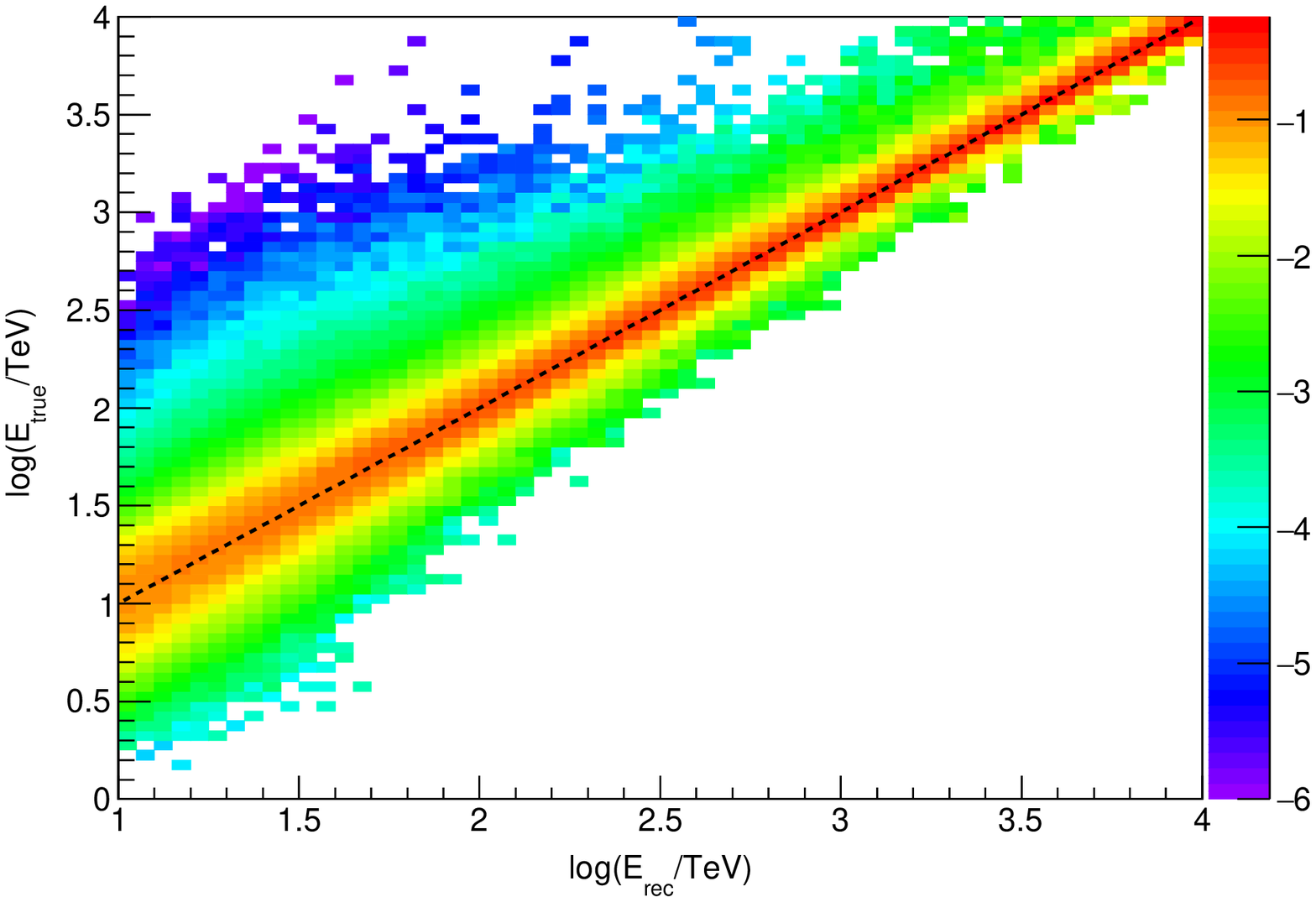}
\includegraphics[width=5.3cm,height=5.3cm]{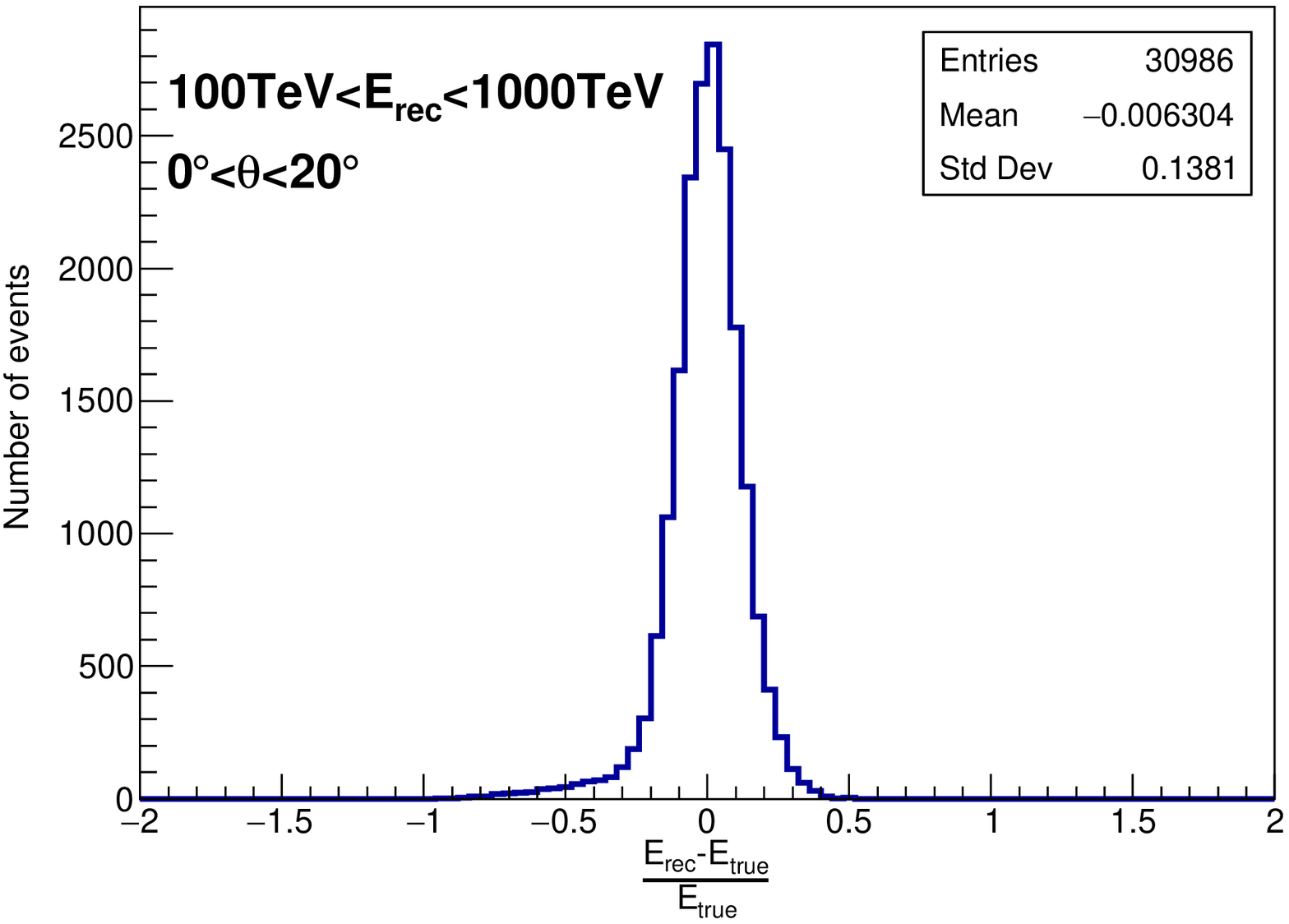}
\includegraphics[width=5.3cm,height=5.3cm]{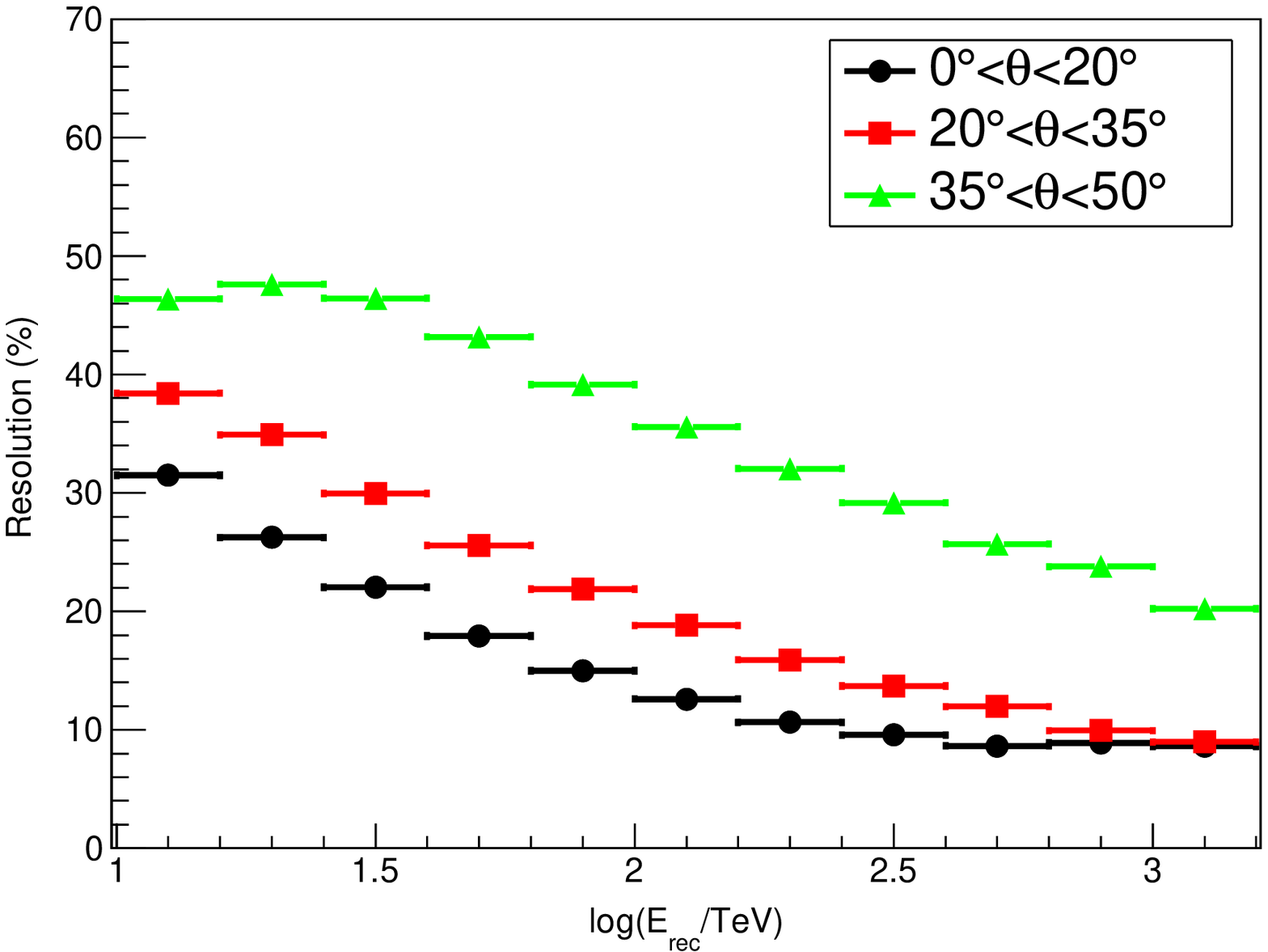}
\figcaption{\label{fig3} The left   panel gives the event-by-event comparison of the primary true energy and the reconstructed energy  for simulated  gamma-ray events over zenith angles   0$^{\circ}$-50$^{\circ}$. The color represents the log probability density    within  each E$_{\rm rec}$ bin.  The dotted line is the identity line. The middle panel shows the energy resolution function of showers in the energy range of  100$-$1000 TeV with zenith angle   0$^{\circ}$-20$^{\circ}$. The right panel is the dependence of energy resolution, defined as the half 68\% width of the resolution function, on each reconstructed energy bin. Three colors indicate the resolutions over   different zenith angle ranges.}
\end{center}
\begin{multicols}{2}

\subsection{Gamma-ray/background discrimination}
Most of the events recorded by KM2A are cosmic ray induced showers, which constitute the chief background for gamma-ray observations.
Considering that gamma-ray induced showers are muon-poor and cosmic rays induced showers are muon-rich, the  ratio between the  measured muons
and electrons is used to discriminate primary gamma-rays from cosmic nuclei.   The ratio is defined as:
\begin{equation}
 R=log(\frac{N_{\rm \mu}+0.0001}{N_{\rm e}})
\end{equation}
where N$_{\rm \mu}$ and N$_{\rm e}$ are defined at the start of Sect. 3, and 0.0001 is used to  show the cases with  N$_{\rm \mu}$=0.
Fig.~\ref{fig7} shows the ratio as a function of the reconstructed energy for gamma-rays and protons.
According to Fig.~\ref{fig7},  the distributions of R for gamma-rays and protons partly overlap at low energies due to wide  N$_{\rm \mu}$ and N$_{\rm e}$  fluctuations.
The separation between gamma-rays and protons become clearer at higher energies.
 For proton showers, the number of electrons detected by EDs is about 10 times   the number of muons detected by MDs.
This factor is about  1000 for gamma-ray showers.
 The shower  illustrated in Fig.~\ref{fig2} is a gamma-ray-like event with N$_{\rm \mu}$=0.

Gamma-ray-like events are selected using simple cuts on the parameter R. These cuts depend on energy and are optimized to maximize the detection
significance (defined by the Li-Ma formula, equation(17) of   \cite{li83}) for a typical Crab-like source.   This optimization consists of a mixture of gamma-ray simulation and real off-source data recorded by KM2A, which are taken to represent the cosmic ray background.
These cuts are shown in  Fig.~\ref{fig7}.
 Fig.~\ref{fig8} shows the
survival fraction for gamma-ray showers (from simulation) along with the measured survival fraction  for the cosmic ray background (from observational data) under these
cuts.  The fraction for gamma-ray showers varies from  48\% to 93\%.   The rejection power of cosmic ray induced showers is better than 4$\times$10$^{3}$    at energies above 100 TeV.

It is worth   noting that these cuts are optimized for point-like sources. The rejection power can be  improved using   stricter cuts. For example, if the survival fraction for gamma-ray showers were restricted to   60\%, the rejection power for cosmic rays would be better than 2$\times$10$^{4}$   at energies above 100 TeV.

\begin{center}
\includegraphics[width=7cm]{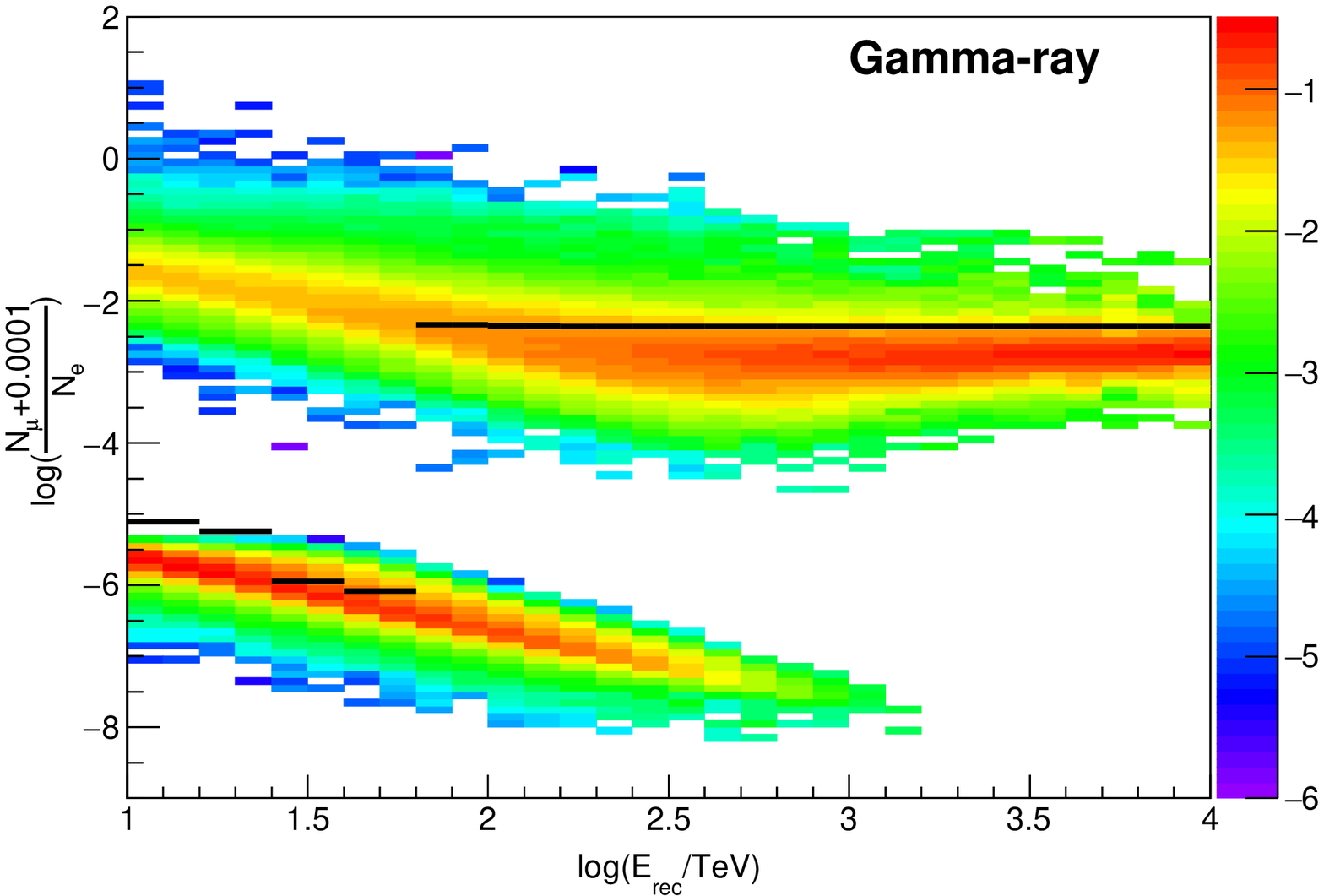}
\includegraphics[width=7cm]{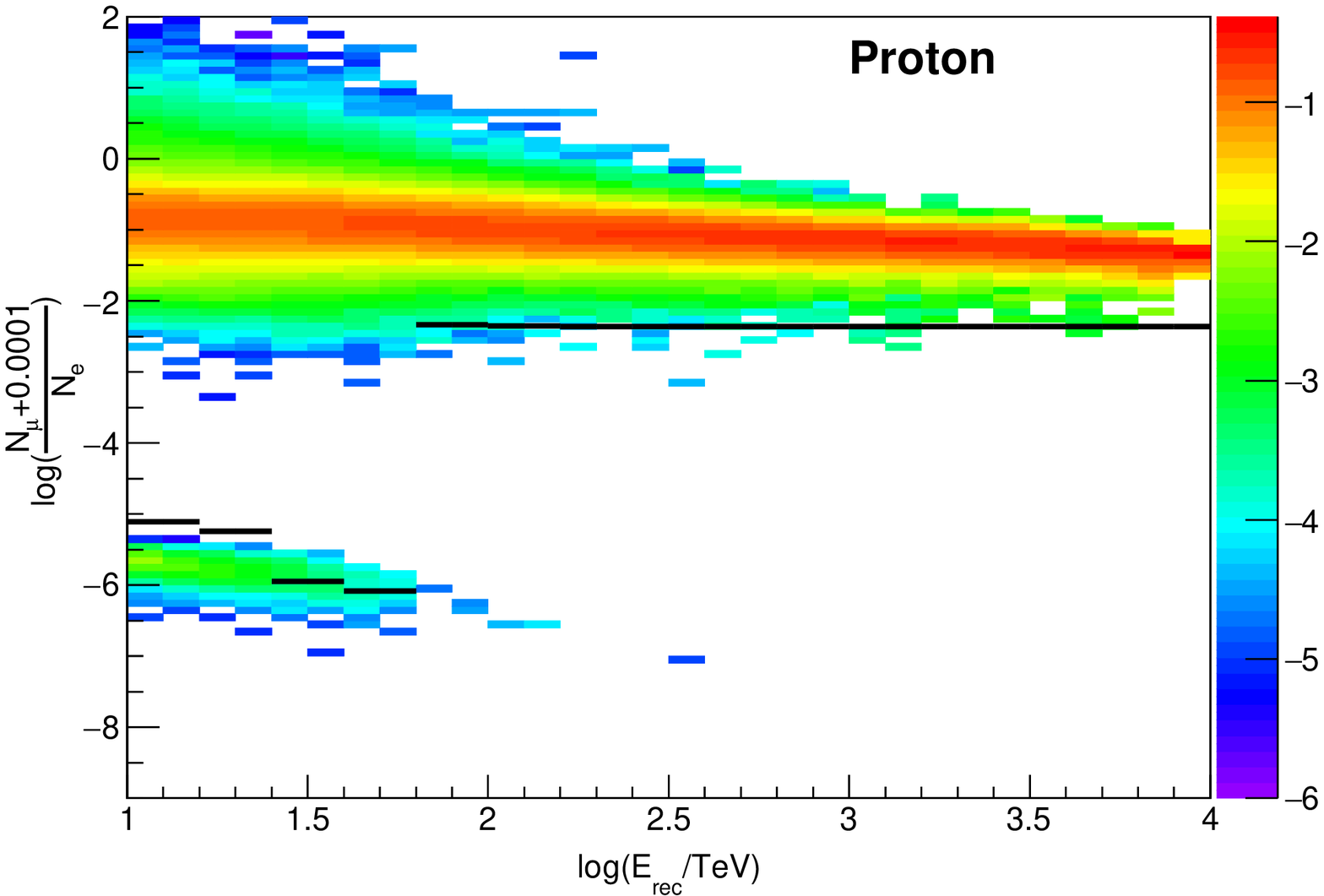}
\figcaption{\label{fig7}Scatter plot of R as defined in equation (6) vs. reconstructed energy using  simulated
gamma-ray-induced (upper panel) and   proton-induced (lower panel) air showers, respectively. The color represents the log probability density     within  each E$_{rec}$ bin. The solid lines indicate the gamma-ray/background discrimination cuts used in this work.}
\end{center}
\begin{center}

\includegraphics[width=7cm]{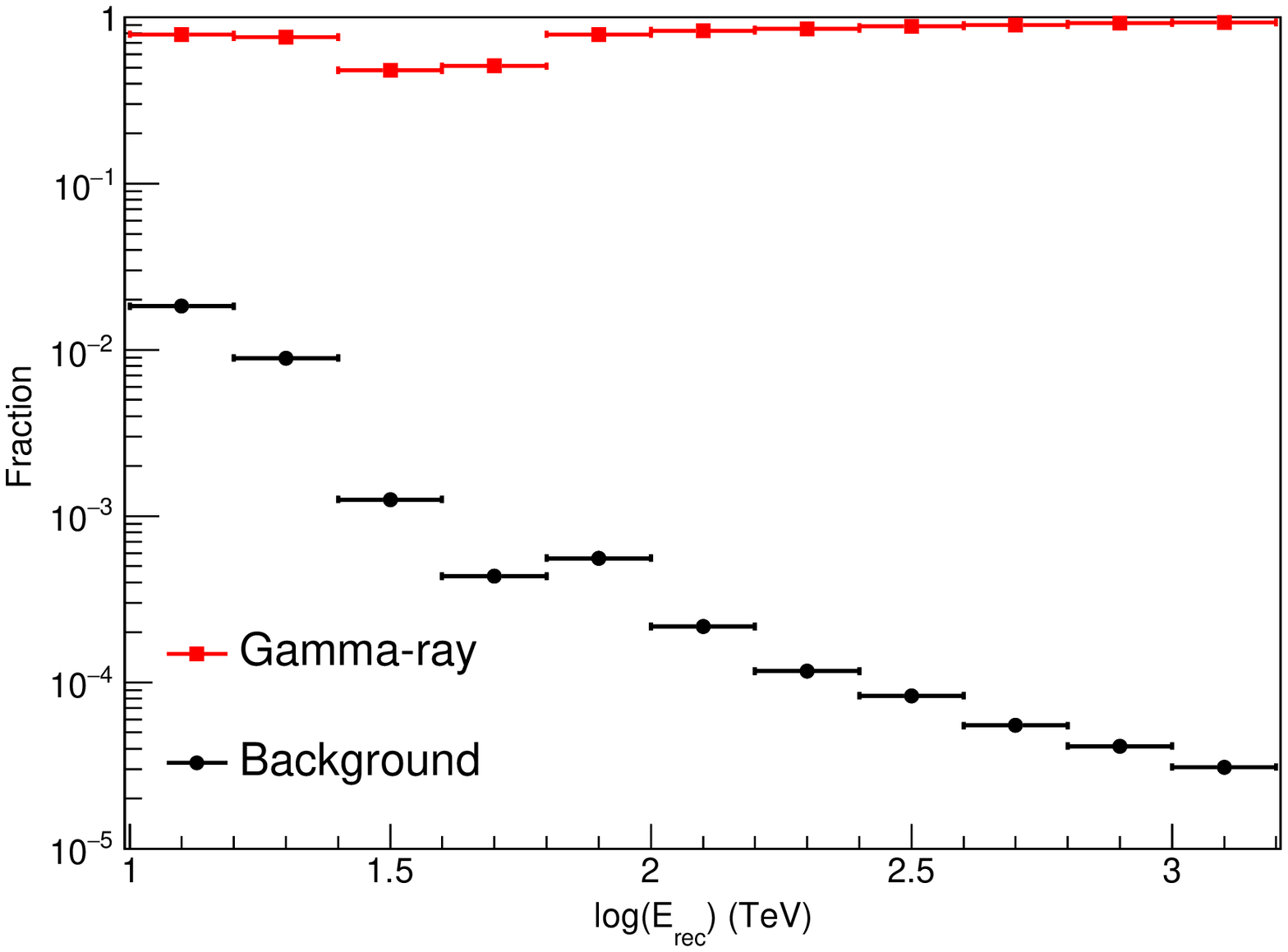}
\figcaption{\label{fig8} The survival fraction of  gamma-ray (according to simulation) and cosmic ray background events (according to observational data) in different energy bins after the  discrimination cuts. }
\end{center}

\subsection{Effective area}
The effective area of the KM2A for detecting gamma-ray showers is calculated using the simulation. It is  energy and zenith angle dependent.
Fig.~\ref{figarea} shows the effective areas  at four zenith angles $\theta=10^{\circ}$, $30^{\circ}$, $40^{\circ}$ and $50^{\circ}$. The  data quality and gamma-ray/background discrimination cuts have been applied here.
The effective area increases with energy and gradually reaches   a constant value at energies above 30 TeV for zenith angles less than $30^{\circ}$.   The effective area is  about 3$\times$10$^{5}$ m$^{2}$ at 20 TeV for a
zenith angle of 10$^{\circ}$.

\begin{center}
\includegraphics[width=7cm]{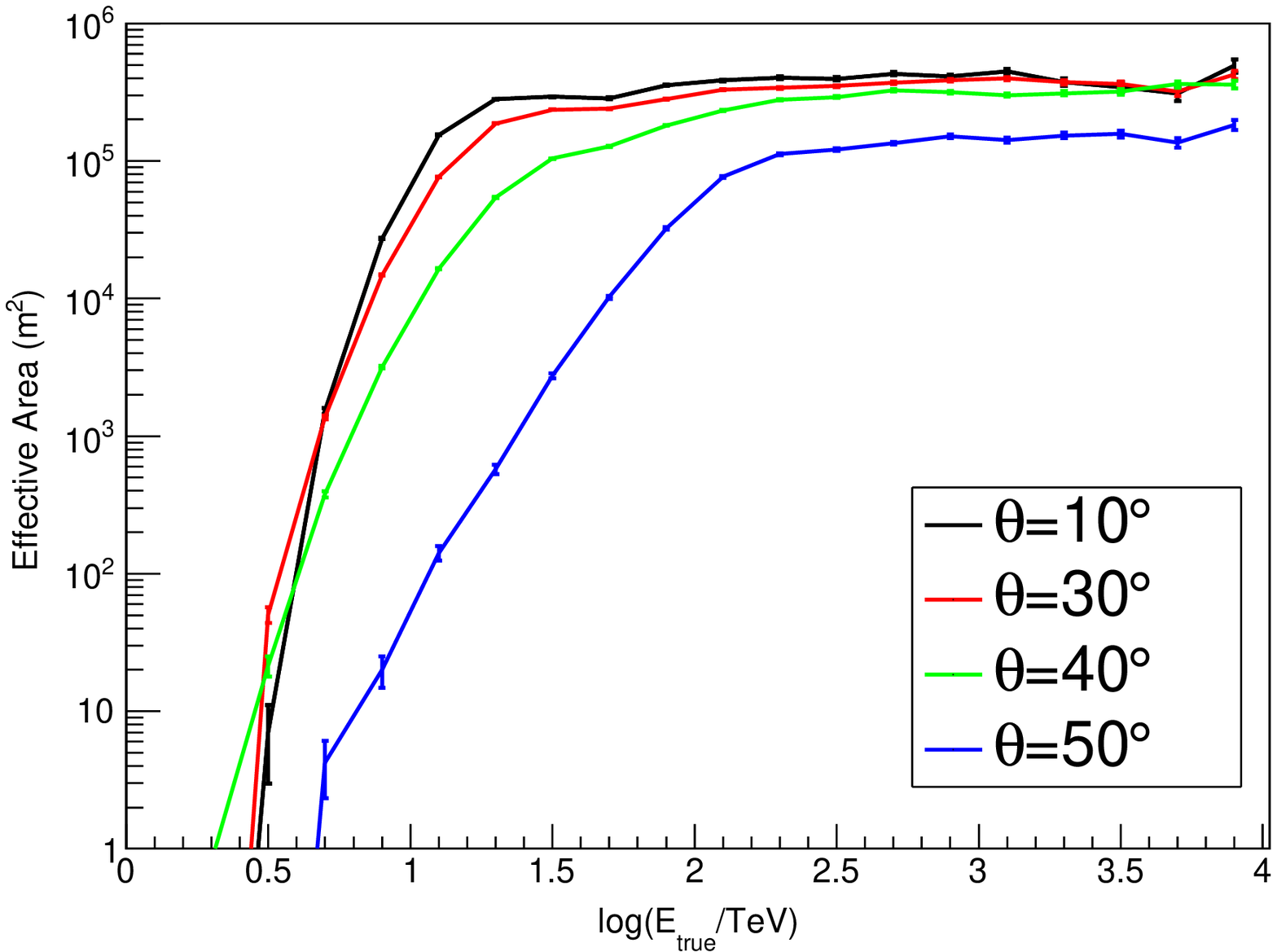}
\figcaption{\label{figarea}The effective area of the KM2A for gamma-ray showers at four zenith angles after applying the data quality and gamma-ray/background discrimination cuts.  }
\end{center}

\section{Analysis and Results}
\subsection{ Background Estimation}
For the analysis presented in this paper, only events with  zenith angles   less than 50 degrees and energies above 10 TeV are used. The  data quality cuts and the gamma-ray/background discrimination cuts discussed in the previous section   are applied.
 The data sets  are divided into five groups per decade according to the reconstructed energy.
For the data set in each group, the sky map in celestial coordinates (right ascension and declination) is divided into a grid of $0.1^{\circ}\times0.1^{\circ}$ pixels
which are filled with the number of the detected events according to their reconstructed arrival direction  (event map).
 To obtain the excess of $\gamma$-induced showers in each pixel, the ``direct integral method'' \cite{fley04}
 is adopted   to estimate the number of cosmic ray background events in the bin.
The ``direct integral method'' uses   events with the same direction in local coordinates but different arrival  time to estimate the background.
In this work, we integrate 24 hours of data to estimate the detector acceptance for different directions. The integral acceptance combined with the event rate is used to estimate the number of background events in each pixel (background map).
This method is   widely used for the ARGO-YBJ \cite{bart13} and HAWC \cite{abey19} experiments.

Then the background map is subtracted from the event map to obtain  the source map which is used to extract the gamma-ray signal from any specific source.  The events in a circular area centered on each pixel within an angular radius  of the KM2 point spread function (PSF)  are summed.
The number of excess events centered on Crab Nebula in each energy bin is used to estimate its gamma-ray spectrum.

\subsection {Data Selection and Significance}
The LHAASO-KM2A data used in this analysis were collected from 27th December 2019  to 28th May 2020. As the beginning of operation, some detectors   still needed debugging during this period. To obtain a reliable data sample, some quality selections have been applied according to the data status. The main selection is to require the number of live EDs $>2100$ and number of live MDs $>500$.   Fig.~\ref{fig90} shows the daily duty cycle after these selections. The average duty cycle is 87.7\% during this period.  The total effective observation time is 136.0 days. With a trigger rate of about 900 Hz, the number of events recorded by KM2A is 1.0$\times$10$^{10}$. After  the data quality cuts and the gamma-ray/background discrimination cuts, the number of events used in
this work is 6$\times$10$^{7}$.
\begin{center}
\includegraphics[width=7cm]{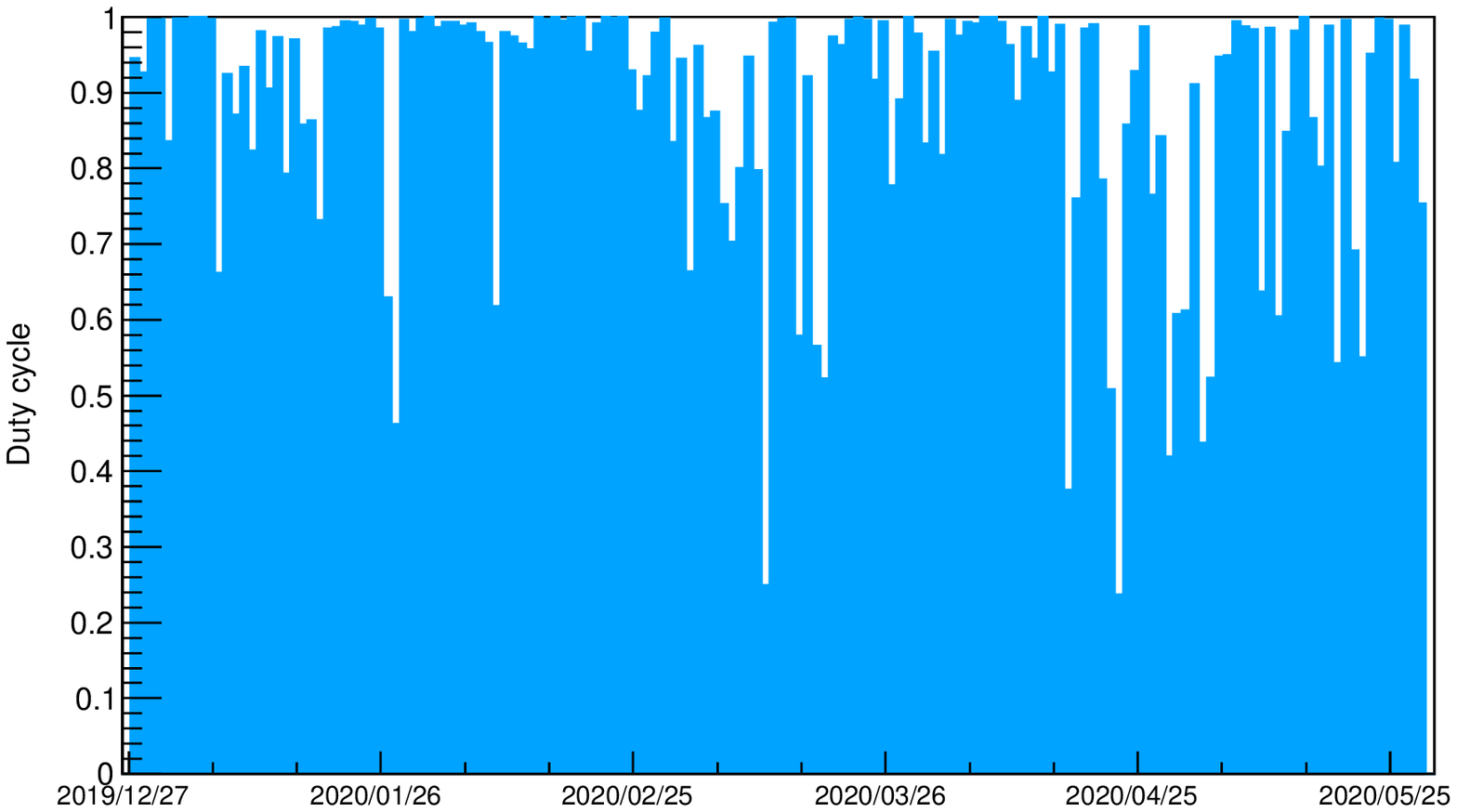}
\figcaption{\label{fig90} The daily duty cycle of 1/2 KM2A operation during the period from  December 2019  to   May 2020.  }
\end{center}

Using these data, the sky  in celestial coordinates  with declination within $-15^{\circ}<$Dec$<75^{\circ}$  is surveyed.
In order to extract a  smooth significance map, the  likelihood method  (see equation 2.5 in \cite{nan17}) is adopted to estimate the significance of the $\gamma$-ray signal. A 2-dimensional Gaussian  is   used to  approximately describe the PSF of the KM2A detector. The width of the Gaussian is set to be  $\sigma_{R}$=$\phi_{68}$/1.51, which is obtained using the simulation sample.
A likelihood ratio test is performed between the background-only model and the one-source model. The test statistic (TS) is used to estimate the significance S=$\sqrt{TS}$. This method is realized by using the MINUIT package.
The pre-trial significance distribution  in the whole sky region at energies above 25 TeV is shown in Fig.~\ref{fig9}.
The distribution  closely follows a standard Gaussian
distribution except for a tail with large positive values, due to
excesses from  gamma-ray emission from the Galactic Plane including the Crab Nebula.  After excluding the Galactic  region with latitude $|b|<12^{\circ}$,   the distribution, with a mean value of -0.05 and $\sigma$= 1.007, closely follows a standard normal
distribution.

\begin{center}
\includegraphics[width=7cm]{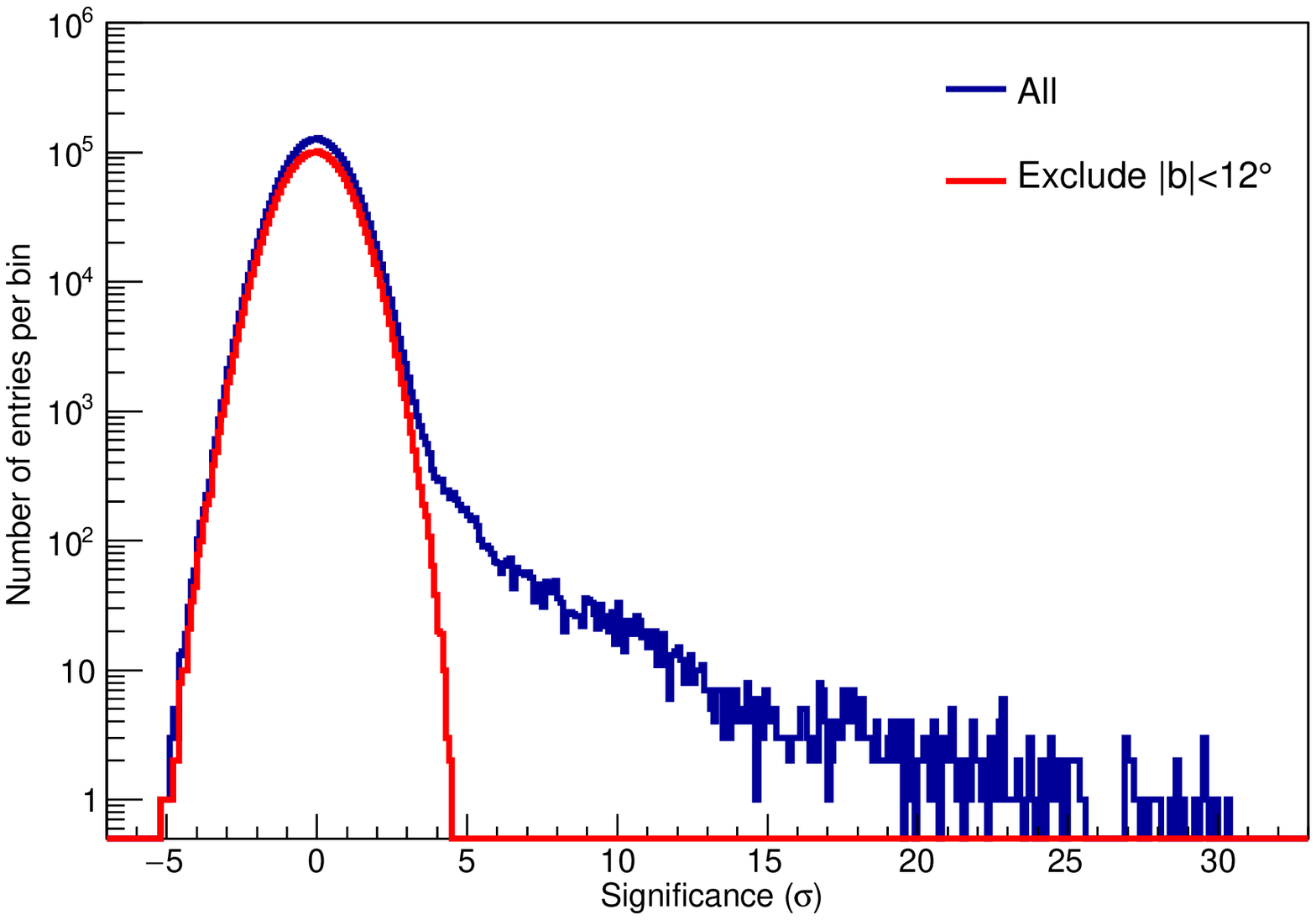}
\figcaption{\label{fig9} Pre-trial significance distribution of events with E$_{\rm rec}>$25 TeV for the whole KM2A sky region (blue) and the portion of the  sky outside the Galactic Plane region with $|b|>12^{\circ}$  (red), which represents the diffuse background events.  }
\end{center}

\begin{figure*}
\begin{center}
\includegraphics[width=6.5in,height=2.1in]{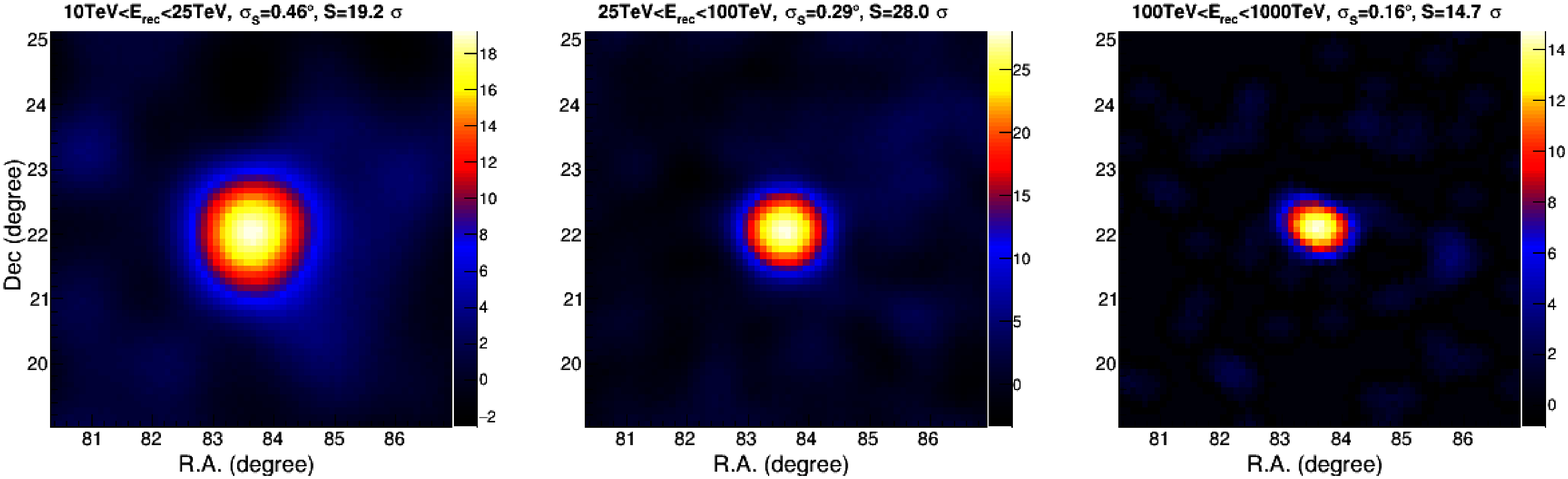}
\caption{Significance maps centered on the Crab Nebula at three energy ranges. $\sigma_{\rm S}$ is   the sigma of the 2-dimension Gaussian taken according to the PSF of KM2A. The color represents the significance.  S is  the maximum value in the map.
}
\label{fig10}
\vspace*{0.5cm}
\end{center}
\end{figure*}

Focusing on the Crab Nebula region, a clear signal   is observed in different energy ranges, i.e., 19.2 $\sigma$ at 10-25 TeV, 28.0 $\sigma$ at 25-100 TeV and 14.7  $\sigma$ at $>$100 TeV  (see Fig.~\ref{fig10}).
 A signal
with such a level of significance allows us to estimate the pointing error of the detector, the angular resolution for gamma-ray showers, and the gamma-ray spectrum from the Crab Nebula.

\subsection{Pointing Accuracy}
To estimate the position of the gamma-ray signal around the Crab Nebula direction at different energy bins, a 2-dimensional Gaussian  is used to fit the event excess map.
The yielded positions in right ascension (R.A.) and declination (Dec) relative to the  known Crab position (R.A.=83.63$^{\circ}$,Dec=22.02$^{\circ}$, J2000.0 epoch) are shown in Fig.~\ref{fig11}. The last energy point in Fig.~\ref{fig11} is obtained using the bins with 100 TeV$<$E$_{\rm rec}<$1 PeV.  When a constant value is used to fit the positions at all energies,   we obtain $\Delta$R.A.=-0.024$^{\circ}\pm$0.016$^{\circ}$,  $\Delta$Dec=0.035$^{\circ}\pm$0.014$^{\circ}$.
\begin{center}
\includegraphics[width=7cm]{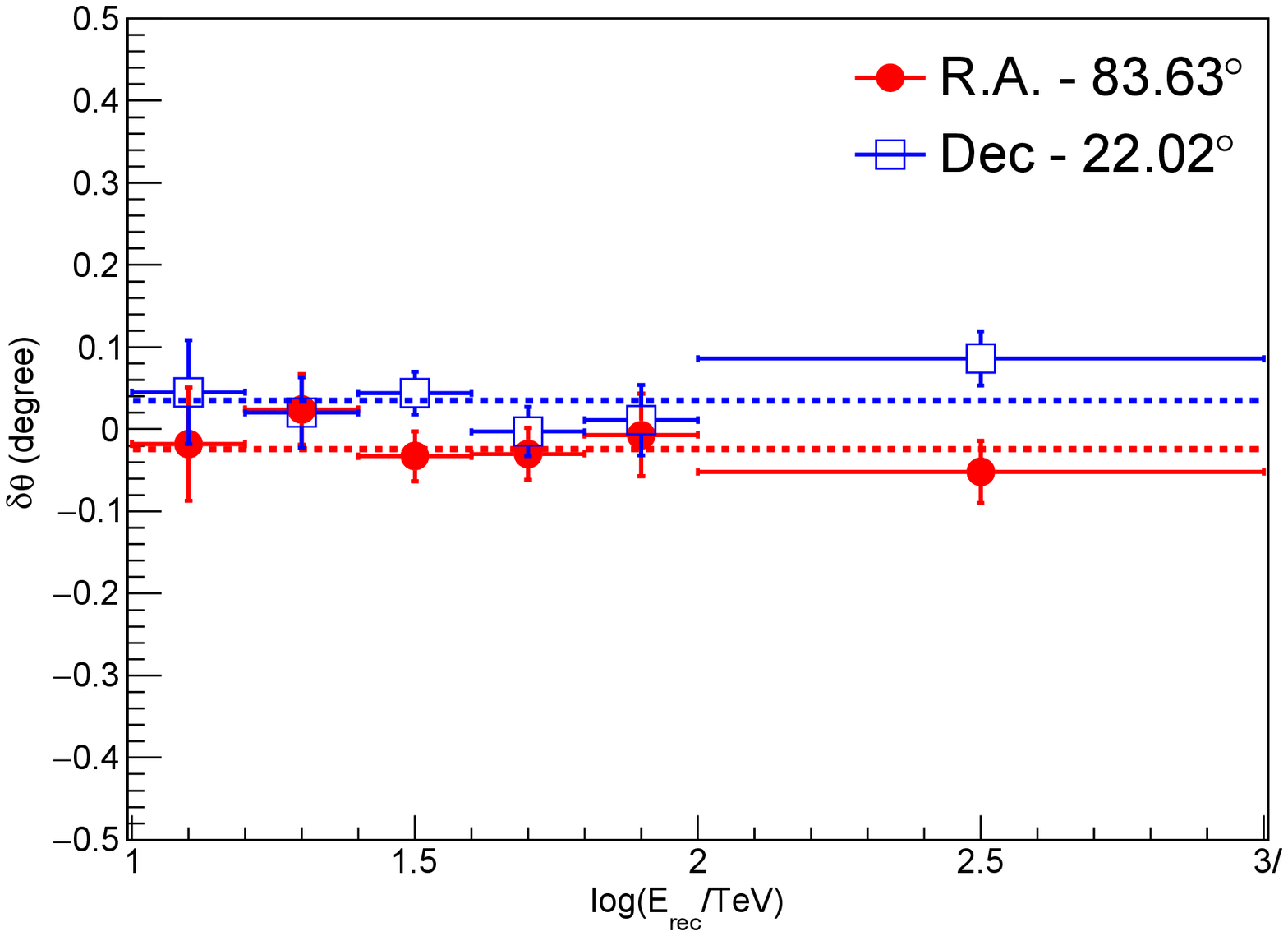}
\figcaption{\label{fig11} The centroid of the significance map around the Crab Nebula  in R.A. and Dec directions  as  a function of  energy. The dashed lines show   constant values that fit the centroid for all energies.   }
\end{center}

The Crab Nebula can be  observed by KM2A for about 7.4 hr per day with a zenith angle less than 50$^{\circ}$, culminating at 7$^{\circ}$. The observation time for zenith angle less than 30$^{\circ}$ is 4.3 hr per day. To check for a possible systematic pointing error at large zenith angles, the observation
of the Crab Nebula at zenith angles higher than 30$^{\circ}$ is analyzed separately. At energies $>$25 TeV, the achieved significance is 12$\sigma$, and the obtained position relative to the known Crab position is $\Delta$R.A.=-0.073$^{\circ}\pm$0.042$^{\circ}$,  $\Delta$Dec=0.074$^{\circ}\pm$0.032$^{\circ}$. This result is roughly consistent with that obtained using all data  within statistical errors.

According to these observations of the Crab Nebula,  the pointing error of KM2A for gamma-ray events can be demonstrated to be  less than 0.1$^{\circ}$.

\subsection{Angular resolution}
\begin{figure*}
\begin{center}
\includegraphics[width=5.3cm,height=5.cm]{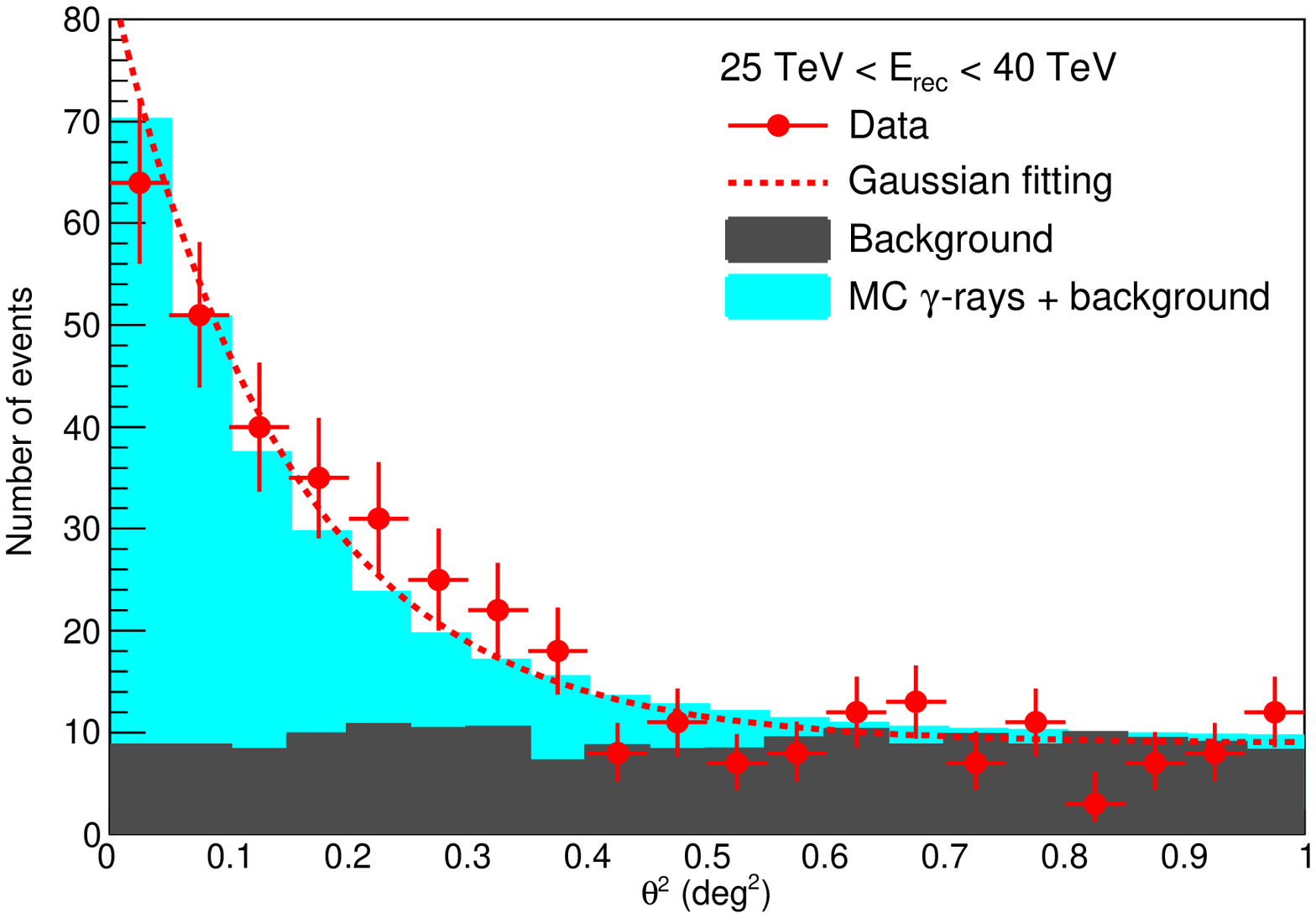}
\includegraphics[width=5.3cm,height=5.cm]{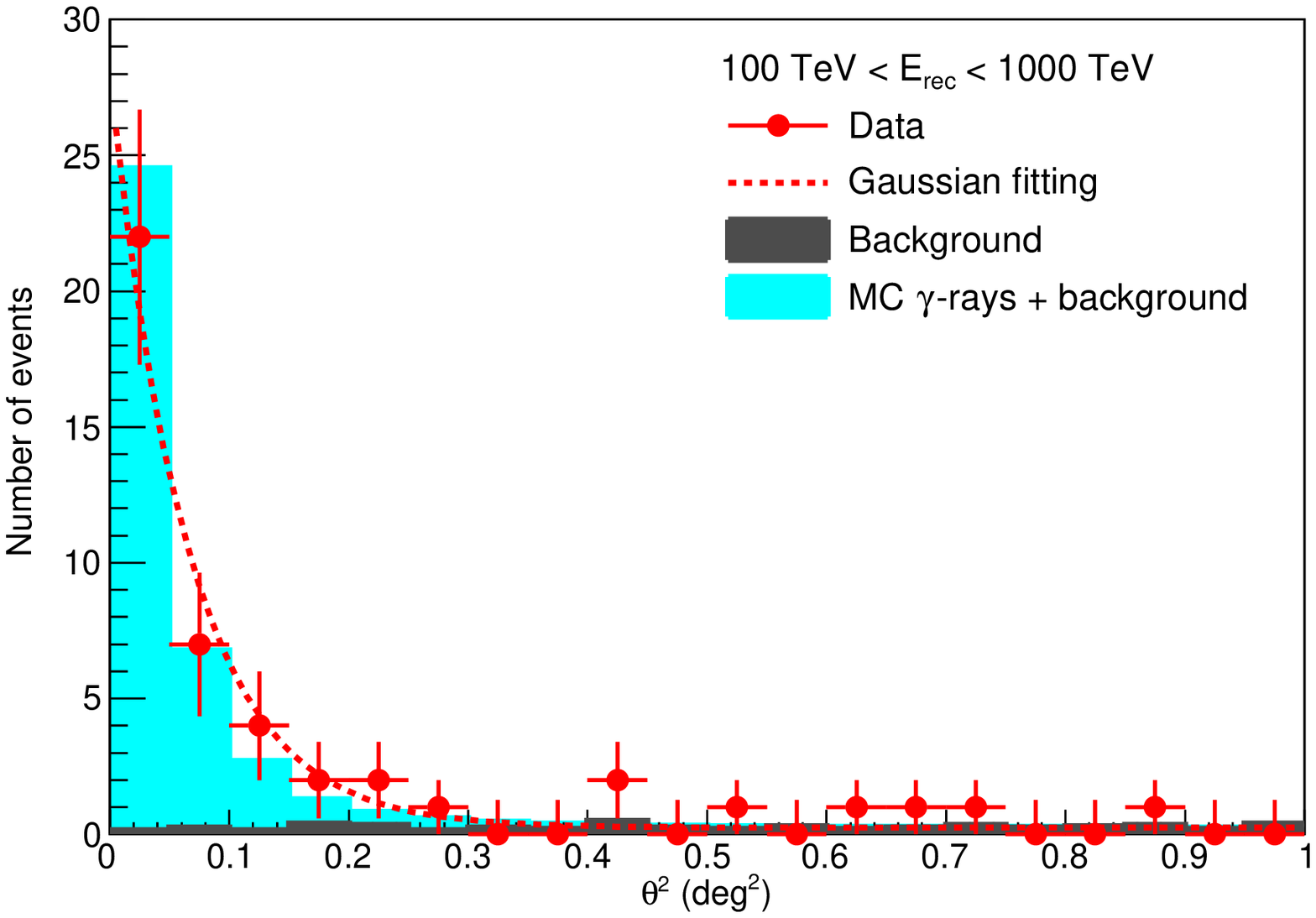}
\includegraphics[width=5.3cm,height=5.cm]{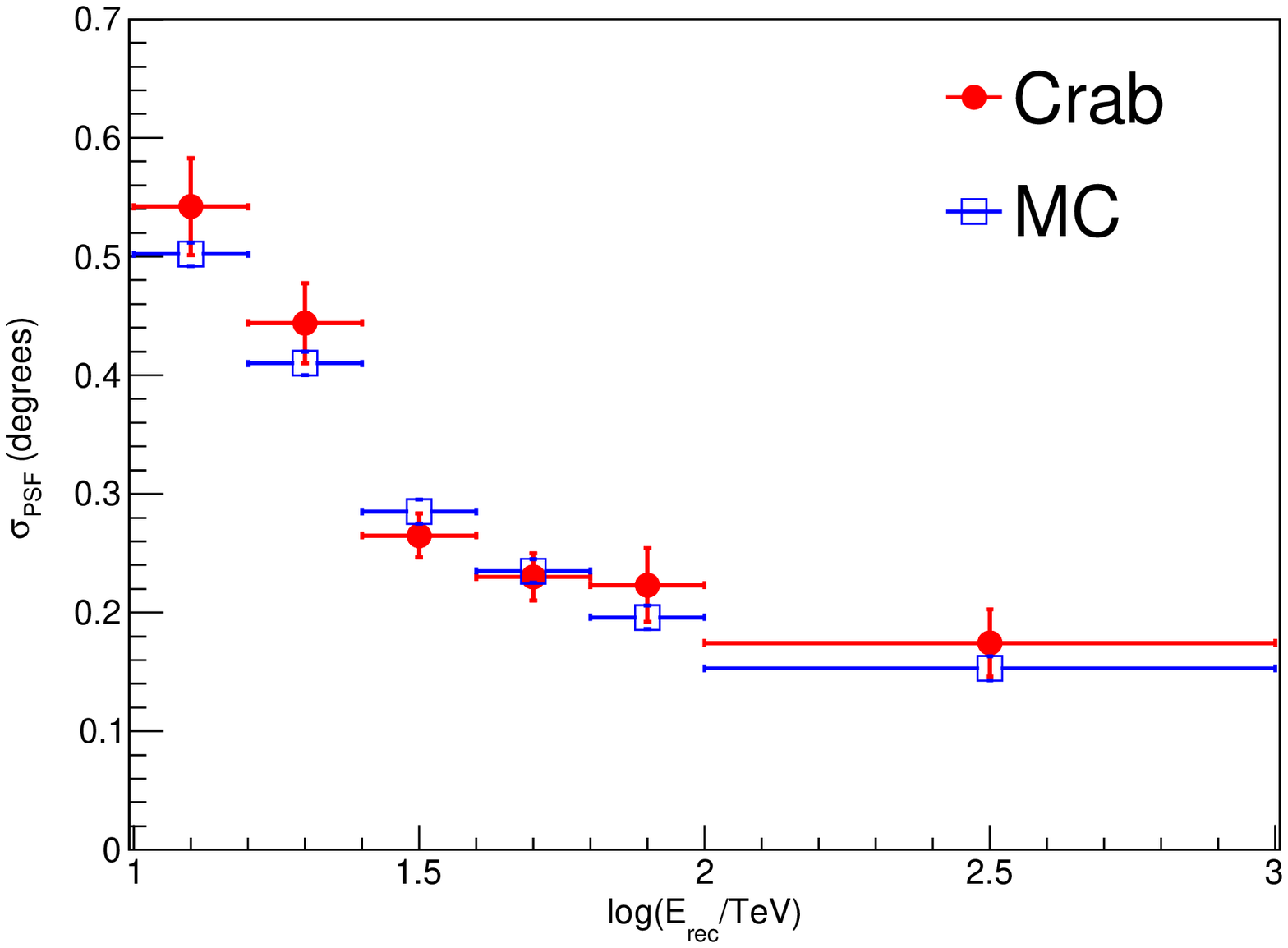}
\figcaption{\label{fig12} Distribution of events as a function of the square
of the    angle to the Crab direction for both experimental data and MC simulation. Two energy ranges, i.e., 25$-$40 TeV (left panel) and 100$-$1000TeV (middle panel) are shown. Right panel is the $\sigma_{PSF}$  as a function of energy.}
\end{center}
\end{figure*}

According to a recent HESS measurement \cite{abda20}, the intrinsic extension of TeV gamma-ray emission from the Crab Nebula is about 0.014$^{\circ}$.
Comparing with  the PSF of the KM2A detector, the intrinsic extension is negligible. Therefore, the angular distribution of gamma-rays detected by KM2A from the Crab Nebula should be mainly due to the detector angular resolution.   Fig.~\ref{fig12} shows the measured angular distribution in KM2A data in two energy ranges.  The solid-angle
density of recorded events  in the vicinity of the Crab Nebula is shown as a function of $\theta^{2}$,  where $\theta$ is the   angle to Crab direction. The distribution is generally consistent with the angular resolution obtained using MC simulations.
For each energy bin,   a Gaussian function is used to fit the angular distribution  shown in the left and middle panels of Fig.~\ref{fig12}. The  resulting  $\sigma_{PSF}$ from Crab data is consistent with      simulations, as shown in the right panel of Fig.~\ref{fig12}.

\subsection{Spectral energy distribution}
The gamma-ray flux from the Crab Nebula is estimated using the   number of excess events (N$_{\rm s}$) and  the corresponding statistical uncertainty ($\sigma_{\rm Ns}$) in each energy bin.
The gamma-ray emission from the Crab Nebula is assumed to follow a power-law spectrum f(E)$=$J$\cdot$E$^{\alpha}$.
The  response of the KM2A detector was  simulated by tracing the trajectory of the Crab Nebula within the FOV of KM2A.
The best-fit values of J and $\alpha$ are obtained by minimizing a $\chi^2$ function for 7 energy bins:
\begin{equation}
\chi^{2}=\sum_{i=1}^{7} \begin{pmatrix} \frac{N_{s_i}-N_{MC_i}(J,\alpha)}{\sigma_{Ns_i}} \end{pmatrix}^{2}
\end{equation}

The resulting differential flux (TeV$^{-1}$ cm$^{-2}$ s$^{-1}$) in
the energy range from 10 TeV to 250 TeV  is:
\begin{equation}
\begin{aligned}
f(E)=(1.13\pm0.05_{stat}\pm0.08_{sys})\times10^{-14}  \\ \begin{pmatrix}\frac{E}{20 TeV}\end{pmatrix}^{-3.09\pm0.06_{stat}\pm0.02_{sys}}
\end{aligned}
\end{equation}

The $\chi^{2}$ of the fit is 1.8 for 5 degrees of freedom, which favors
a pure power-law description of the spectrum.
The SED is shown  in Fig.~\ref{fig13} and is also listed in Table 1.
The SED obtained in this work is in agreement with previous  observations by other detectors, such as HEGRA \cite{ahar04},  HAWC \cite{abey19} and Tibet
AS-$\gamma$ \cite{amen19}.

\end{multicols}
\begin{center}
\tabcaption{ \label{tab2} Energy and differential flux   as
shown in Fig.~\ref{fig13}}
\footnotesize
\begin{tabular*}{90mm}{c|cccc}
\toprule
 log(E$_{\rm rec}/TeV$) &  E$_{middle}$ & N$_{on}$ & N$_{b}$   & Differential Flux  \\
                     &  (TeV)          &        &         & (TeV$^{-1}$ cm$^{-2}$ s$^{-1}$))\\
\hline
 [1.0, 1.2] &  12.6 & 10810 & 9620 &  (4.52$\pm$0.40)$\times$10$^{-14}$   \\
\hline
[1.2, 1.4]  &  20.0 & 2513  & 1902&  (1.13$\pm$0.09)$\times$10$^{-14}$   \\
\hline
[1.4, 1.6]  &  31.6 & 294   &  81  & (2.98$\pm$0.24)$\times$10$^{-15}$  \\
\hline
[1.6, 1.8] &  50.1  & 91    &  9.3   & (6.64$\pm$0.78)$\times$10$^{-16}$        \\
\hline
[1.8, 2.0] &  79.4  & 47    &  4.0   & (1.43$\pm$0.23)$\times$10$^{-16}$  \\
\hline
[2.0, 2.2] &  126   & 21    &  0.50   & (4.05$\pm$0.91)$\times$10$^{-17}$ \\
\hline
[2.2, 2.4] &  200   & 7    &  0.11   & (8.00$_{-3.19}^{+3.84}$)$\times$10$^{-18}$ \\
\bottomrule
\end{tabular*}
\vspace{0mm}
\end{center}
\vspace{0mm}

\begin{multicols}{2}

\begin{center}
\includegraphics[width=7cm]{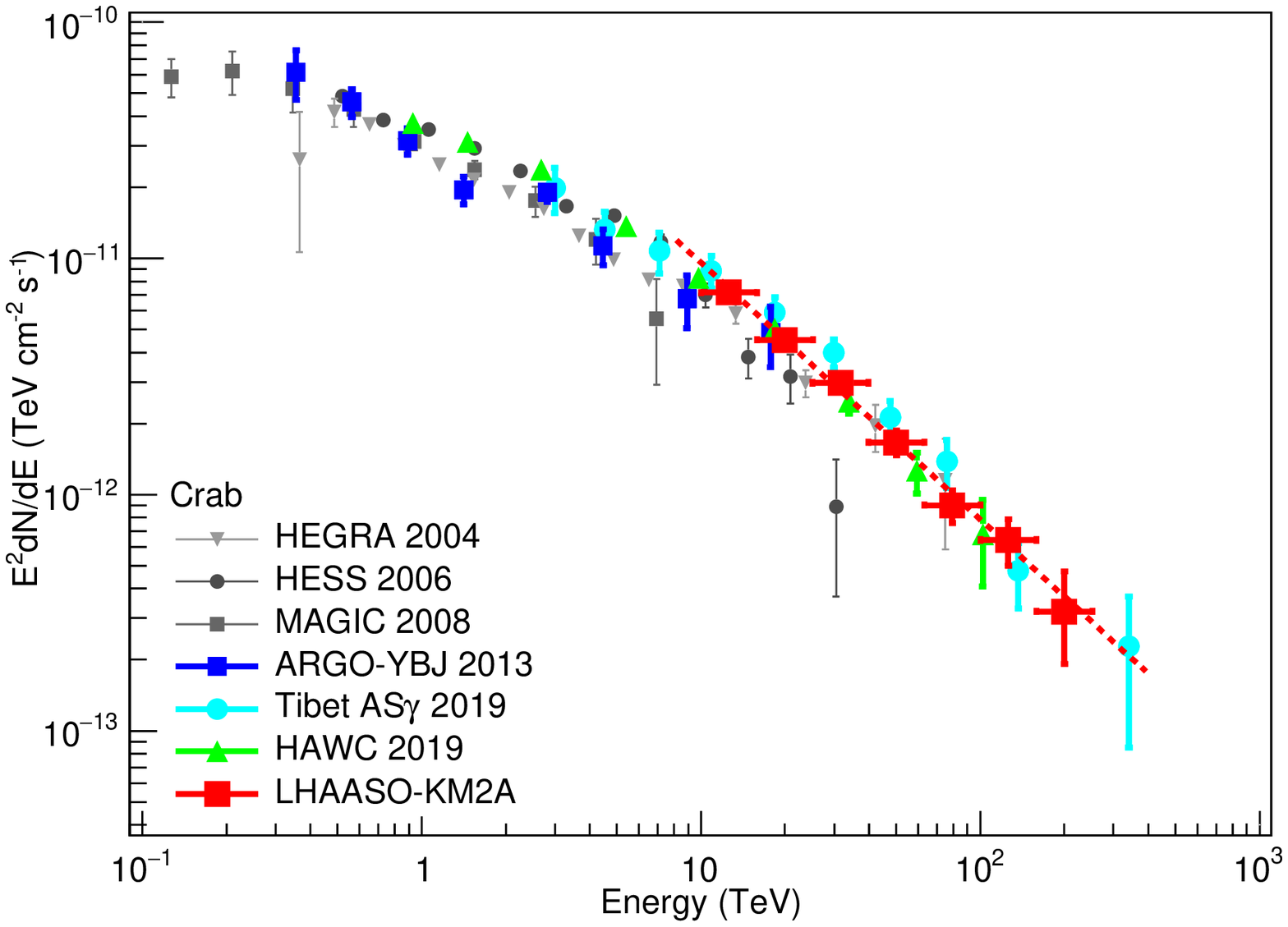}
\figcaption{\label{fig13} The spectrum of the Crab Nebula measured by KM2A in red together with the spectra measured by other experiments in various colors as indicated in the legend.    The dotted line indicates the best fitting result using a power-law  function. References for other experiments are:
HEGRA \cite{ahar04}, HESS \cite{ahar06}, MAGIC \cite{albe08}, ARGO-YBJ \cite{bart13},  HAWC \cite{abey19}, Tibet AS-$\gamma$ \cite{amen19}.  }
\end{center}

\subsection {Systematic Uncertainties}
The systematic errors affecting the SED have been investigated by studying the variation of  the Crab Nebula spectrum under various assumptions.
During the period of interest, about a few percent of detector units was under debugging. The number of operating  units varied with time. A typical layout is taken into account in the detector simulation   to mimic the status of the array.  The uncertainty is estimated by using different configurations in the detector simulation. The variation of detector number  affects  the gamma-ray/background separation, while the impact on gamma-rays is weaker than   on the background. The maximum variation in flux introduced by detector layout is less than 2\%.
The main systematic error  comes from the atmospheric model used in the Monte Carlo simulations.
The atmospheric density profile in reality always deviates from the model provided  in \cite{heck98}  due  to the seasonal and daily   changes.  According to the variation of event rate during the operational  period, the total systematic uncertainty   is estimated to be  7\% on the flux and 0.02 on the spectral index.

\section{Summary}
Using the first five months of data from the KM2A half-array, a standard candle at very high energy ---  Crab Nebula --- is observed to investigate  the detector performance and corresponding data analysis pipeline for gamma-rays. The  statistical significance of the gamma-ray signal from Crab Nebula is  28.0 $\sigma$ at 25-100 TeV and 14.7  $\sigma$ at $>$100 TeV. The gamma-ray angular distributions around the source are fairly consistent with the point spread function  obtained by simulations. According to measurement of the centroids of the significance maps of the Crab Nebula  at different energies, the  pointing error of KM2A is found to be  less than 0.1$^{\circ}$.  The  spectrum from 10 TeV to 250 TeV is well fitted with a   power-law function with a spectral index of $3.09\pm0.06_{stat}\pm0.02_{sys}$. This result is quite consistent with previous measurements by other experiments. The overall systematic error of KM2A on spectral measurement is  estimated to be  7\% in flux and   0.02 in spectral index.

The pipeline of KM2A data analysis  presented in this work is not   specifically designed for the  Crab Nebula but  also generally useful  for surveying the whole sky in the range of  declination from -15$^{\circ}$ to 75$^{\circ}$ and  the corresponding measurements for the source morphology and energy spectrum.
This opens a new window of gamma-ray astronomy above 0.1 PeV. A new era of  ultrahigh-energy gamma-ray astronomy is foreseen the fruitful with fundamental discoveries.

\vspace{3mm}
\acknowledgments{
This work is supported in
China by National Key R\&D program of China under the grants 2018YFA0404201,
2018YFA0404202,  2018YFA0404203,  by NSFC (No.12022502, No.11905227, No.11635011, No.11761141001, No.U1931112, No.11775131,  No.U1931201, No.11905043), and  in Thailand by RTA6280002 from Thailand Science Research and Innovation.
The authors would like to thank all staff members who work at the LHAASO site above 4400 meters above   sea level year-round to maintain the detector and keep the electrical power supply and other components of the experiment operating smoothly. We are grateful to the Chengdu Management Committee
of Tianfu New Area for their constant financial support  of research  with LHAASO data.
}

\end{multicols}

\vspace{-1mm}
\centerline{\rule{80mm}{0.1pt}}
\vspace{2mm}

\begin{multicols}{2}

\end{multicols}

\clearpage
\end{CJK*}
\end{document}